\documentclass[aip,jcp,reprint,amsmath,amssymb]{revtex4-2}
\usepackage[T1]{fontenc}
\usepackage{graphicx}
\usepackage{dcolumn}
\usepackage{bm}
\usepackage{braket}
\usepackage[colorlinks=true,citecolor=blue,linkcolor=blue,urlcolor=black]{hyperref}

\begin{document}

\title{Coherent and dissipative energy transfer
in driven open quantum systems: a driven quantum master equation approach}

\author{Junran Kong}
\author{Yitian Chen}
\author{Huan Liu}
\author{Liwei Duan}\email{duanlw@zjnu.edu.cn}
\author{Chen Wang}\email{wangchen@zjnu.cn}

\affiliation{Department of Physics, Zhejiang Normal University, Jinhua 321004, China}

\begin{abstract}

Understanding how coherent driving and quantum dissipation cooperatively control nonequilibrium energy transfer is challenging in driven open quantum systems. Here, we develop a driven quantum master equation (dQME) that uses the eigenstates of the full driven Hamiltonian in the rotating frame and retains the driving phase in the system-reservoir interactions. This approach includes driving effects in both coherent evolution and dissipative exchange processes. 
We apply this approach to a
coupled-qubit charger-battery system and dissipative energy transfer through one and two driven qubits.
For a driven coupled-qubit battery, weak driving produces three peaks in stored energy near the central and two side resonances,
whereas strong driving merges these peaks into a broad central peak. 
Ergotropy and charging efficiency are suppressed at the weak-driving resonances as battery coherence decreases. 
The central suppression of charging efficiency persists under strong driving. 
For dissipative energy transfer, 
the driven single-qubit model reveals three energy exchange channels set by the driving frequency and the dressed-state energy gap. 
Interqubit coupling produces additional current resonances, while strong driving broadens the frequency range of enhanced energy currents. 
We hope that the dQME may provide a common framework for understanding coherent energy charging and dissipative energy transfer.

\end{abstract}

\maketitle

\section{Introduction}

Quantum control and measurement of driven quantum matter are of fundamental importance for quantum science~\cite{Sieberer2025rmp,wiseman2011book}, enabling the modulation of quantum responses via time-dependent coherent and incoherent external fields~\cite{segal2008prl,bai2021aipx,wang2022fop,huang2026prl}.
Active driving protocols enable a range of quantum technologies, including laser driving~\cite{benlloch2014prl,Munoz2021prl,mohsen2026prl}, energy pumping~\cite{ren2010prl,Nathan2021prl,Magazzu2021pra},
and quantum thermodynamic machines~\cite{Brandner2020prl,wang2024prl,lu2025prb,micadei2019nc}.
External modulation  enables high-precision control for advanced quantum technologies, e.g., quantum computing and sensing~\cite{wu2025npj,Engelhardt2025pra,Menta2025prr}.

The study of driven open quantum systems focuses on active nonequilibrium processes~\cite{oka2019arcmp,wang2022fop}, representing a significant departure from passive counterparts.
When considering the interplay between the external driving and quantum dissipation, these driven open systems show nontrivial nonequilibrium steady states beyond traditional equilibrium states~\cite{wang2024prl,mohsen2025prb,saha2026prb}. 
This intriguing interplay deepens the understanding of energy and matter flows in quantum systems, 
yielding new nonequilibrium physics, such as quantum time crystals~\cite{gong2018prl,li2024prl}, exceptional points~\cite{miri2019science,zhang2024prl,jiang2025cpl}, and nonequilibrium phase transitions~\cite{zhu2020prl,mohsen2023prb,beaulieu2025nc}. 

Recently, the quantum battery, a device able to coherently store and supply useful energy at the quantum level, has attracted increasing attention~\cite{alicki2013pre,zhang2019pre,campaioli2024rmp}.
In particular, the superextensive power scaling was explored by Ferraro \emph{et al.} via collective quantum light-matter interactions~\cite{Ferraro2018prl,Crescente2020prb,dou2022prb,dou2022pra}. 
In practice, a quantum battery inevitably interacts with its thermal environment.
The dissipative quantum battery therefore emerges as a generic active open device for work extraction~\cite{barra2019prl,ferraro2026nrp},
which can be strengthened by quantum coherence~\cite{mayo2022pra}, quantum collisions~\cite{seah2021prl}, and the quantum Mpemba effect~\cite{medina2025prl}.
In investigations of indirect quantum charger-battery interactions,
the wireless quantum battery was discovered by Song \emph{et al.} to perform long-range charging, immune to environmental noise~\cite{song2024prl,yang2025pra,hu2025pra}.

To actively excite a quantum charger subject to dissipation,
a coherent driving field is generally included in a dissipative quantum battery, yielding a driven-dissipative quantum battery. Efficient coherent energy transfer can be realized via
quadratic-photon driving~\cite{downing2023cp}, nonreciprocal engineering~\cite{ahmadi2024prl,khan2026prxe},
pure dephasing~\cite{shastri2025npjqi},
and the self-discharging effect~\cite{song2025prl}.
In contrast to closed quantum batteries, which focus on short-time dynamics to optimize stored energy and extracted work,
 driven-dissipative quantum batteries alternatively center on steady-state charging energy and efficiency~\cite{campaioli2024rmp,downing2023cp}, where quantum dissipation affects the features of quantum batteries.
Methodologically, the Lindblad master equation is generally used to study charging performance with a weak external driving field and weak coupling between internal components~\cite{downing2023cp,ahmadi2024prl}.
For driving fields that become strong compared with dissipation, distinct dressed pictures of the quantum master equation have been proposed to investigate dissipative dynamics~\cite{wang2025jpcl,boite2017pra,boite2020aqt}.
However, the simultaneous inclusion of driving information in the incoherent transition processes of the quantum master equation is still missing.

Meanwhile, dissipative energy transfer under external driving in nonequilibrium open systems has attracted sustained interest,
with studies addressing adiabatic and nonadiabatic protocols~\cite{wang2022fop,wang2024prl,Takahashi2020prl}, stochastic pumping~\cite{segal2008prl}, as well as Floquet engineering~\cite{vahid2024jcp,vahid2025prb}.
Specifically, for time-periodic driving in the adiabatic limit, the geometric-phase-induced incoherent pumping was discovered~\cite{wang2022fop,ren2010prl}, which significantly modifies energy exchange processes.
For nonadiabatic driving, the higher-order geometric terms should be considered to correctly capture the energy transfer and fluctuations~\cite{Takahashi2020prl,Funo2020prl}.
To describe dissipative transfer,
one generally requires sophisticated frameworks, e.g.,
nonequilibrium Green functions~\cite{Maciejko2006prb,vahid2024jcp}, the hierarchical equations of motion~\cite{Sakurai2013jpsj},
Floquet theory~\cite{grifoni1998rmp,Engelhardt2024prr},
and first-principles schemes~\cite{zhang2013prb}. 
Quantum master equations can also be combined with full counting statistics~\cite{esposito2009rmp} to track the complete distribution of transferred charge and energy.
In particular, for finite coupling between internal subsystems,
the time-dependent quantum master equation (tQME)~\cite{boite2017pra,boite2020aqt} was proposed by Boite \emph{et al.} for coherent driving that is weak compared with the quantum dissipation strength,
where the time-dependent driving was directly included in the unitary evolution part of tQME.
For coherent modulation beyond the weak-driving limit, 
although Floquet theory can handle the driven system Hamiltonian, the computational cost of expanding Floquet space poses a challenge.
Hence, a concise treatment of time-dependent driving and quantum dissipation within the quantum master equation framework is still far from satisfactory.

Here, we propose a driven quantum master equation (dQME) in the rotating frame to study quantum energy transfer in general open quantum systems under external periodic driving.
The main points are as follows:
(1) dQME retains the driving phase in dissipative transition processes, which modifies microscopic transition rates relative to the traditional tQME.
(2) Steady-state charging performance is investigated in a coupled-qubit system from weak to strong driving amplitudes, in which three resonances merge into a single broadened resonance.
The underlying microscopic transfer mechanisms are demonstrated.
(3) Dissipative energy transfer is analyzed in a driven spin-boson model, where the influence of the three-channel structure on energy currents is analytically explored.
Interqubit coupling induces additional resonance structure.
This paper is structured as follows: In Sec.~\ref{sec:doqs}, we introduce the general driven open quantum system and the von Neumann equation in the rotating frame. In Sec.~\ref{sec:qme}, we
derive the dQME and describe its distinction from the traditional tQME.
We highlight the crucial roles of driving frequency and amplitude in optimizing energy currents. 
In Sec.~\ref{sec:result1}, we study the coherent energy transfer in a driven-dissipative quantum battery system.
In Sec.~\ref{sec:result2}, we study the incoherent energy transfer in driven nonequilibrium spin systems.
We present a summary in Sec.~\ref{sec:conclusion}.

\begin{figure*}[t]
\centering
\includegraphics[width=17cm,height=21.1cm,keepaspectratio]{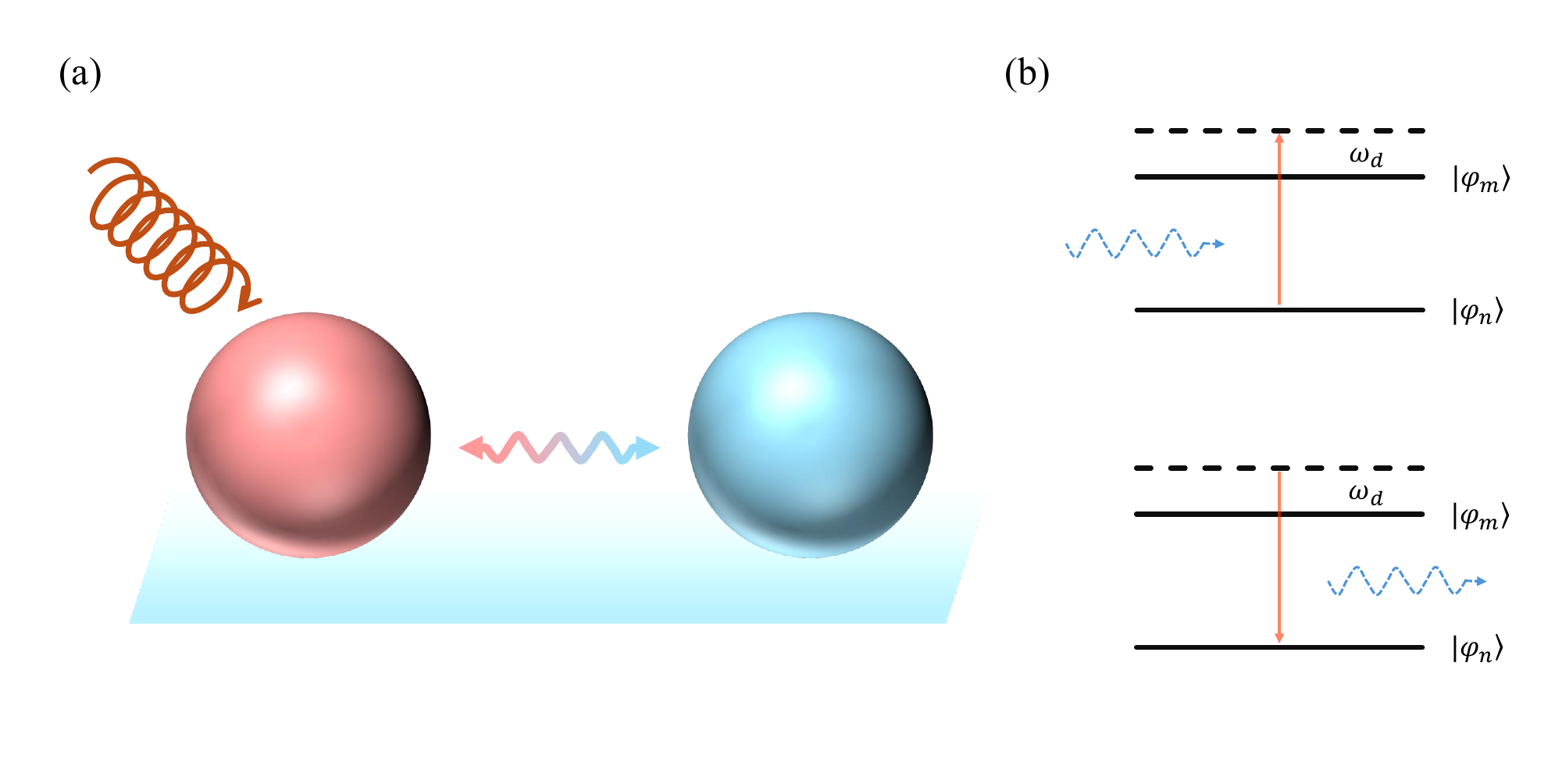}
\caption{
(a) Schematic of a driven open quantum system $\hat{H}_{\rm tot}$ in Eq.~(\ref{drivenH0}),
where the red and blue spheres denote a general coupled quantum system,
the brown spiral line with an arrow represents the external driving field applied to the quantum system,
and the bottom blue pane represents the
bosonic thermal environment.
(b) Two typical drive-assisted upward (downward) incoherent transitions from $|\phi_{n(m)}{\rangle}$ to
$|\phi_{m(n)}{\rangle}$, described by Eqs.~(\ref{Dmu+}) and (\ref{Dmu-}).
}\label{fig1}
\end{figure*}

\section{Driven open quantum system}\label{sec:doqs}

\subsection{Coupled quantum system}
We include an external driving field in a generic coupled quantum system for quantum transfer, with each component individually interacting with a bosonic thermal environment, as illustrated in Fig.~\ref{fig1}(a).
The general driven-dissipative quantum system can be described as (setting $\hbar=1$)
\begin{align}\label{drivenH0}
\hat{H}_{\textrm{tot}}(t)=\hat{H}_{\textrm{DS}}(t)+\sum_{\mu=l,r}(\hat{H}^\mu_{b}+\hat{V}_\mu).
\end{align}
The driven central quantum system is expressed as
\begin{align}\label{drivenH1}
\hat{H}_{\textrm{DS}}(t)&=\varepsilon\sum_{\mu=l,r}\hat{A}^\dag_\mu\hat{A}_\mu+\lambda(\hat{A}^\dag_l\hat{A}_r+\hat{A}^\dag_r\hat{A}_l)\nonumber\\
&-{D}(e^{-i\omega_dt}\hat{A}^\dag_l+e^{i\omega_dt}\hat{A}_l),    
\end{align}
where $\hat{A}^\dag_\mu~(\hat{A}_\mu)$ denotes the creation (annihilation) operator of a general field (e.g., fermionic, spin, or bosonic) with the occupation energy $\varepsilon$,
$\lambda$ is the coupling strength between system components,
and
$D$ and $\omega_d$ are the driving amplitude and frequency.
The $\mu$th bosonic thermal reservoir is described as
$\hat{H}^\mu_{b}=\sum_k\omega_{k,\mu}\hat{b}^\dag_{k,\mu}\hat{b}_{k,\mu}$,
where $\hat{b}^\dag_{k,\mu}~(\hat{b}_{k,\mu})$ creates (annihilates) one boson in the reservoir with the frequency $\omega_{k,\mu}$.
The linear system-bath interactions are given by
\begin{align}\label{Vmu}
\hat{V}_\mu=\sum_k(g_{k,\mu}\hat{b}^\dag_{k,\mu}\hat{A}_\mu+g^{*}_{k,\mu}\hat{b}_{k,\mu}\hat{A}^{\dag}_\mu),
\end{align}
with $g_{k,\mu}$ the coupling strength.

Then, we study the dynamics of the driven-dissipative quantum system in the rotating frame.
Specifically, we apply the rotation operator $\hat{R}(t)=\exp(-i\omega_dt\sum_{\mu}\hat{A}^\dag_\mu\hat{A}_\mu)$
to the total system density operator $\hat{\rho}_\textrm{R}(t)=\hat{R}^\dag(t)\hat{\rho}_\textrm{tot}(t)\hat{R}(t)$,
leading to the von Neumann equation in the rotating frame
\begin{align}\label{EqR1}
\frac{d}{dt}\hat{\rho}_\textrm{R}(t)=-i[\hat{H}_\textrm{R}(t),\hat{\rho}_\textrm{R}(t)].
\end{align}
The rotated Hamiltonian is described by
$\hat{H}_\textrm{R}(t)=\hat{H}+\hat{H}_{b}+\hat{V}_\textrm{R}(t)$.
The rotated quantum system is expressed as
\begin{align}
    \hat{H}_{s,\rm R}&=\Delta\sum_{\mu=l,r}\hat{A}^\dag_\mu\hat{A}_\mu+\lambda(\hat{A}^\dag_l\hat{A}_r+\hat{A}^\dag_r\hat{A}_l)\nonumber\\
    &-{D}(\hat{A}^\dag_l+\hat{A}_l),~\label{RHs1}
\end{align}
where the detuning is $\Delta=\varepsilon-\omega_d$.
We note that after the rotating transformation, the central quantum system in Eq.~(\ref{drivenH1}) becomes time-independent.
The system-bath interactions, however, become time-dependent, i.e.,
\begin{align}
\hat{V}_\textrm{R}(t)=\sum_{k,\mu}(g_{k,\mu}e^{-i\omega_dt}\hat{b}^\dag_{k,\mu}\hat{A}_\mu+g^{*}_{k,\mu}e^{i\omega_dt}\hat{b}_{k,\mu}\hat{A}^{\dag}_\mu).\label{vR}    
\end{align}

\subsection{Driven quantum master equation}\label{sec:qme}

Starting from the von Neumann equation $\frac{d}{dt}\hat{\rho}_\textrm{R}(t)=-i[\hat{H}_\textrm{R}(t),\hat{\rho}_\textrm{R}(t)]$ at Eq.~(\ref{EqR1}) in the rotating frame, we find that the dynamics of the total density operator in the interaction picture can be reexpressed as
$ \frac{d}{dt}\hat{\rho}^\textrm{R}_{I}(t)=-i[\hat{V}^\textrm{R}_{I}(t),\hat{\rho}^\textrm{R}_{I}(0)]
 -\int^t_0d\tau[\hat{V}^\textrm{R}_{I}(t),[\hat{V}^\textrm{R}_{I}(\tau),\hat{\rho}^\textrm{R}_{I}(\tau)]]$,
where $\hat{\rho}^\textrm{R}_{I}(t)=\exp(i\hat{H}_0t)\hat{\rho}^\textrm{R}(t)\exp(-i\hat{H}_0t)$, with $\hat{H}_0=\hat{H}_{s,\rm R}+\hat{H}_{b}$.
Here we assume the quantum system is simultaneously coupled to several bosonic thermal reservoirs.
Assuming weak system-bath interactions and fast relaxation of thermal baths, we include the Born approximation $\hat{\rho}^\textrm{R}_{I}(t){\approx}\hat{\rho}^\textrm{R}_{s,I}(t)
{\otimes}\hat{\rho}_{b}$,
where the equilibrium density operator of the reservoirs is
$\rho_{b}=\exp(-\beta\hat{H}_{b})/\textrm{Tr}[\exp(-\beta\hat{H}_{b})]$. 
By tracing out the degrees of freedom of the bosonic baths and considering the linear system-reservoir interactions $[\hat{V}_{\rm R},\hat{\rho}^\textrm{R}_{I}(0)]=0$,
we find the dynamics of the reduced system density operator
$d\hat{\rho}^\textrm{R}_{s,I}(t)/dt= -\int^t_0d\tau\textrm{Tr}_b\{
 [\hat{V}^\textrm{R}_{I}(t),[\hat{V}^\textrm{R}_{I}(\tau),\hat{\rho}^\textrm{R}_{I}(\tau)]]\}$.
Then, we consider the Markov approximation $\hat{\rho}^\textrm{R}_{I}(\tau){\approx}\hat{\rho}^\textrm{R}_{I}(t)$ to simplify the dissipative dynamics as
\begin{align}\label{append:qme1}
 \frac{d}{dt}\hat{\rho}^\textrm{R}_{s,I}(t)= -\int^t_0d\tau\textrm{Tr}_b\{
 [\hat{V}^\textrm{R}_{I}(t),[\hat{V}^\textrm{R}_{I}(\tau),\hat{\rho}^\textrm{R}_{I}(t)]]\}.
 \end{align}
One crucial step here is to find the explicit expressions of $\hat{V}^\textrm{R}_{I}(t)=\sum_{\mu}e^{i\hat{H}_0t}\hat{V}^\textrm{R}_\mu(t)e^{-i\hat{H}_0t}$,
which is expressed by
\begin{align}\label{VRI}
    \hat{V}^\textrm{R}_{I,\mu}(t)=\sum_k[g_{k,\mu}e^{i(\omega_{k,\mu}-\omega_d)t}\hat{b}^\dag_{k,\mu}\hat{A}_{I,\mu}(t)
+\rm{H.c.}],
\end{align}
where the system operators are given by $\hat{A}_{I,\mu}(t)=e^{i\hat{H}t}\hat{A}_{\mu}e^{-i\hat{H}t}$.
We note that the system-reservoir couplings in the interaction picture exhibit additional time-dependent behaviors, carrying the driving phase $\exp({\pm}i\omega_dt)$.
This driving information embedded in the system-reservoir interactions may dramatically affect the dynamical and steady-state behaviors of open quantum systems.

We then return to the Schr\"odinger picture from Eq.~(\ref{append:qme1}) and replace the variable $(t-\tau)$ by $\tau$.
The driven quantum master equation can be reexpressed as
\begin{align}\label{qme1}
	\frac{d}{dt}\hat{\rho}_{s,\rm R}(t)&=-i[\hat{H}_{s,\rm R},\hat{\rho}_{s,\rm R}]\nonumber\\
	&+\sum_\mu\bigl\{[\hat{D}^\dag_{\mu,+}\hat{\rho}_{s,\rm R},\hat{A}_\mu]\nonumber\\
&\qquad
	+[\hat{D}_{\mu,-}\hat{\rho}_{s,\rm R},\hat{A}^\dag_\mu]+\mathrm{H.c.}\bigr\},
\end{align}
where the modified system operators in the Markov approximation are described as
\begin{subequations}
	\begin{align}		\hat{D}_{\mu,+}&=\frac{1}{2}\sum_{m,m^\prime}\gamma_\mu(\omega_d+E_{m^\prime{m}})n_\mu(\omega_d+E_{m^\prime{m}})\nonumber\\
&\qquad\times A^\mu_{mm^\prime}|\phi_m{\rangle}{\langle}\phi_{m^\prime}|,~\label{Dmu+}\\
    \hat{D}_{\mu,-}&=\frac{1}{2}\sum_{m,m^\prime}\gamma_\mu(\omega_d+E_{m^\prime{m}})[1+n_\mu(\omega_d+E_{m^\prime{m}})]\nonumber\\
&\qquad\times A^\mu_{mm^\prime}|\phi_m{\rangle}{\langle}\phi_{m^\prime}|,~\label{Dmu-}		
	\end{align}
\end{subequations}
where  the spectral function is described as $\gamma(\omega)_\mu=2\pi\sum_k|g_{k,\mu}|^2\delta(\omega-\omega_{k,\mu})$, the Bose-Einstein distribution function is
$n_{\mu}(\omega)=1/[\exp(\omega/k_{\textrm{B}}T_\mu)-1]$,
with $k_\textrm{B}$ the Boltzmann constant and $T_\mu$ the $\mu$th bosonic reservoir's temperature,
the transition coefficient is specified as
$\hat{A}^\mu_{mm^\prime}={\langle}\phi_m|\hat{A}_\mu|\phi_{m^\prime}{\rangle}$,
the eigenstates are given by $\hat{H}_{\rm S,R}|\phi_n{\rangle}=E_n|\phi_n{\rangle}$, and the energy gap is $E_{mn}=E_m-E_n$.
From the system operators at Eqs.~(\ref{Dmu+}) and (\ref{Dmu-}), we find that
the incoherent transition processes describe downward (upward) transitions from $|\phi_{m^\prime}{\rangle}~(|\phi_{m}{\rangle})$ to 
$|\phi_{m}{\rangle}~(|\phi_{m^\prime}{\rangle})$ assisted by the quanta $\omega_d$ of the driving field, as shown in Fig.~\ref{fig1}(b).
In this study, we use an Ohmic spectral function~\cite{leggett1987rmp}, i.e.
$\gamma_\mu(\omega)=\alpha_\mu\theta(\omega)(\omega/\varepsilon)\exp(-\omega/\omega_c)$, with $\alpha_\mu$ the dissipation strength,
$\omega_c$ the cutoff frequency of thermal baths,
and $\theta(\omega)$ the Heaviside step function,
i.e.,
$\theta(\omega>0)=1$
and
$\theta(\omega{\leq}0)=0$.
Generalization to other spectral functions is straightforward.

Interestingly, it is found that the local Gibbs detailed balance relation is generally broken, which is characterized as $\Gamma^\mu_{+}(\omega_d+E_{mn})/\Gamma^\mu_{-}(\omega_d+E_{mn}){\neq}\exp(-E_{mn}/k_{\textrm{B}}T_\mu)$,
due to the microscopic inclusion of the external driving frequency.
In the limit $\omega_d{\rightarrow}0$, the local detailed balance relation is restored
$\Gamma^\mu_{+}(E_{mn})/\Gamma^\mu_{-}(E_{mn}){=}\exp(-E_{mn}/k_{\textrm{B}}T_\mu)$.
In the opposite limit $\omega_d{\gg}\{E_{m^{\prime}m}\}$, there exists another straightforward relation
$\Gamma^\mu_{+}(\omega_d)/\Gamma^\mu_{-}(\omega_d){=}\exp(-\omega_d/k_{\textrm{B}}T_\mu)$, which implies that the incoherent transitions are synchronized by the external driving.

We also note that in some works the rotation operator is modified by
$\hat{R}(t)=\exp[-i\omega_dt\sum_{\mu}(\hat{A}^\dag_\mu\hat{A}_\mu+\sum_{k}\hat{b}^\dag_{k,\mu}\hat{b}_{k,\mu})]$~\cite{yan2014pra}.
Accordingly, the rotated system Hamiltonian is identical to Eq.~(\ref{RHs1}).
The thermal reservoirs are expressed as $\hat{H}^\prime_{b,\mu}=\sum_k(\omega_k-\omega_d)\hat{b}^\dag_{k,\mu}\hat{b}_{k,\mu}$.
The system-reservoir interactions are given by
$\hat{V}_\textrm{R}=\sum_{k,\mu}(g_{k,\mu}\hat{b}^\dag_{k,\mu}\hat{A}_\mu+g^{*}_{k,\mu}\hat{b}_{k,\mu}\hat{A}^{\dag}_\mu)$, which become time-independent.
However, in the interaction picture
such interactions also include the driving phase, i.e.,
$\hat{V}^\textrm{R}_{I,\mu}(t)=\sum_k[g_{k,\mu}e^{i(\omega_{k,\mu}-\omega_d)t}\hat{b}^\dag_{k,\mu}\hat{A}_{I,\mu}(t)
+g^*_{k,\mu}e^{-i(\omega_{k,\mu}-\omega_d)t}\hat{b}_{k,\mu}\hat{A}^\dag_{I,\mu}(t)]$,
as also shown in Eq.~(\ref{VRI}).
Moreover, under the Born approximation the equilibrium reservoir density operator is invariant under the rotating transformation,
expressed as 
$\hat{\rho}_b=\Pi_{\mu}[\exp(-\hat{H}^\mu_{b}/k_\textrm{B}T)/\mathcal{Z}_\mu]$, with $\hat{H}^\mu_{b}$ in Eq.~(\ref{drivenH0}).
Hence, the identical driven quantum master equation at Eq.~(\ref{qme1}) is recovered.

\subsection{Time-dependent quantum master equation}

In previous works, the time-dependent quantum master equation (tQME) describes the coherent time-dependent driving processes by including the external driving term in the unitary evolution of the master equation~\cite{boite2017pra,boite2020aqt}.
In the dressed picture, the driving term is specified as $\hat{V}(t)=-D(e^{-i\omega_dt}\hat{A}^\dag_l+e^{i\omega_dt}\hat{A}_l)$.
Thus, tQME is approximately described as
\begin{align}
\frac{d}{dt}\hat{\rho}_{s}(t)&=
i[\hat{\rho}_{s}(t),\hat{H}_s-D(e^{-i\omega_dt}\hat{A}^\dag_l+e^{i\omega_dt}\hat{A}_l)]\nonumber\\
&+\sum_{\mu,m,m^\prime}
\bigl\{\Gamma^\mu_{-}(\Delta_{m^{\prime}m})\mathcal{L}[|\psi_m{\rangle}{\langle}\psi_{m^\prime}|]\hat{\rho}_{s}(t)\nonumber\\
&\qquad+\Gamma^\mu_{+}(\Delta_{m^{\prime}m})\mathcal{L}[|\psi_{m^\prime}{\rangle}{\langle}\psi_{m}|]\hat{\rho}_{s}(t)\},   ~\label{dme0}
\end{align}
where $\hat{\rho}_{s}(t)$ denotes the density operator of the reduced quantum system,
\begin{align}
\hat{H}_{s}=\varepsilon\sum_{\mu=l,r}\hat{A}^\dag_\mu\hat{A}_\mu-\lambda(\hat{A}^\dag_l\hat{A}_r+\hat{A}^\dag_r\hat{A}_l)~\label{Hstatic}    
\end{align}
is the static system Hamiltonian component, as shown in Eq.~(\ref{drivenH1}),
$|\psi_m{\rangle}$ is the eigenstate of $\hat{H}_{s}$ with the corresponding eigenvalue $\Delta_{m}$,
the incoherent transition rates at Eq.~(\ref{dme0}) are approximately treated as
$\Gamma^\mu_{\pm}(\Delta_{m^{\prime}m})=\pm\gamma_\mu(\Delta_{m^{\prime}m})n_\mu({\pm}\Delta_{m^{\prime}m})|{\langle}\psi_m|\hat{A}_\mu|\psi_{m^\prime}{\rangle}|^2$, with $n_\mu(-\omega)=-[1+n_\mu(\omega)]$,
the dissipator is specified as
\begin{align}
    \mathcal{L}[\hat{O}]\hat{\rho}_s=(2\hat{O}\hat{\rho}_s\hat{O}^\dag-\hat{O}^\dag\hat{O}\hat{\rho}_s-\hat{\rho}_s\hat{O}^\dag\hat{O})/2.
\end{align}
We note that the dressed basis $|\psi_m{\rangle}$ is limited to the eigenvectors of $\hat{H}_{\rm S}$, excluding the driving terms.

We point out the main features of dQME:
(1) The dQME uses the eigenstates of $\hat{H}_{s,\rm R}$, the rotated central Hamiltonian, which includes the rotated static component and the driving term $-D(\hat{A}^\dag_l+\hat{A}_l)$.
For tQME, the traditional selection of the dressed basis is instead based on $\hat{H}_{s}$.
The dressed pictures used in these two methods are generally distinct and coincide only in the weak-driving limit.
Hence, we provide a concise and interpretable master equation framework to investigate driven-dissipative dynamics and energy transfer.
(2) The driving frequency $\omega_d$ is naturally embedded into the incoherent transition operators at Eqs.~(\ref{Dmu+}-\ref{Dmu-}), due to the additional phase $e^{\pm{i\omega_d}t}$ in the system-bath interactions in the interaction picture at Eq.~(\ref{VRI}).
In contrast, this information is missing from the tQME. 
As $\omega_d$ increases, it may dramatically affect dynamical and steady-state behaviors of the driven-dissipative quantum systems.
The dQME also has limitations:
(1) dQME is restricted to the monochromatic driving protocol.
Hence, a two-tone driving protocol is beyond the scope of this study.
(2) Weak system-bath interactions and rapid relaxation of the thermal reservoirs are required for the Born-Markov approximation.

\section{Coherent energy charging}\label{sec:result1}

\subsection{Performance metrics}
A quantum energy battery generally describes a bipartite physical system harnessing
genuine quantum features, e.g., quantum coherence and quantum uncertainty, to store energy.
The battery is supplied by an external quantum source, i.e., a quantum charger~\cite{alicki2013pre,zhang2019pre,campaioli2024rmp}.
Several representative quantities characterize the charging performance of a quantum battery.
In particular, for a dissipative quantum battery, 
researchers focus on the charging energy, the charging ergotropy, and the charging efficiency after long-time evolution, because
the short-time dynamics is mainly determined by the unitary evolution of the system Hamiltonian $\hat{H}_{\rm DS}(t)$.
Specifically, for a bipartite quantum system at Eq.~(\ref{drivenH1}),
the $l$-th component represents the quantum charger,
and the $r$-th component describes the quantum battery.
Then, the charging energy is the energy stored in the  battery at time $\tau$
\begin{align}
    E_{\rm B}(\tau)=\textrm{Tr}_r\{\hat{H}_{r}\hat{\rho}_r(\tau)\},
\end{align}
where the battery Hamiltonian is $\hat{H}_{r}=\varepsilon\hat{A}^\dag_r\hat{A}_r$, the reduced density operator of the battery is obtained as $\hat{\rho}_r(\tau)=\textrm{Tr}_{l}\{\hat{\rho}_s(\tau)\}$, with $\hat{\rho}_s$ the density operator of the coupled charger-battery system.
Similarly, the energy of the charger can also be described as
\begin{align}
    E_{\rm C}(\tau)=\textrm{Tr}_l\{\hat{H}_{l}\hat{\rho}_l(\tau)\},
\end{align}
where
$\hat{H}_{l}=\varepsilon\hat{A}^\dag_l\hat{A}_l$ and the reduced density operator of the charger is $\hat{\rho}_l(\tau)=\textrm{Tr}_{r}\{\hat{\rho}_s(\tau)\}$.
The charging ergotropy~\cite{abn2004el,wang2026cpl}, i.e., the extractable work from the battery system, 
is expressed as
\begin{align}\label{erg1}
    \mathcal{E}_{\rm B}(\tau)= E_{\rm B}(\tau)-\min_{\hat{U}_{\rm B}}\textrm{Tr}_r\{\hat{H}_{r}\hat{U}_{\rm B}\hat{\rho}_r(\tau)\hat{U}^\dag_{\rm B}\},
\end{align}
where the minimization at Eq.~(\ref{erg1}) is performed
over all possible unitary operators $\hat{U}_{\rm B}$.
When no additional work can be extracted from the battery, 
the minimization is reached.
The corresponding state is called the passive state $\hat{\sigma}_r$,
i.e., $\textrm{Tr}_r\{\hat{H}_{r}\hat{\sigma}_r\}{\le}\textrm{Tr}_r\{\hat{H}_{r}\hat{U}_{\rm B}\hat{\rho}_r(\tau)\hat{U}^\dag_{\rm B}\}$.
One can straightforwardly express the battery Hamiltonian as
$\hat{H}_{r}=\sum_nE_n|E_n{\rangle}{\langle}E_n|$
with the eigenenergy $E_n$ in ascending order,
and diagonalize the density operator as
$\hat{\rho}_r=\sum_mr_m|r_m{\rangle}{\langle}r_m|$
with $r_m$ in descending order.
Then, the passive state can be constructed as
$\hat{\sigma}_r=\sum_nr_n|E_n{\rangle}{\langle}E_n|$.
Consequently, the charging efficiency is expressed as
\begin{align}
    \eta_{\rm B}(\tau)=\frac{\mathcal{E}_{\rm B}(\tau)}{E_{\rm B}(\tau)},
\end{align}
which characterizes the energy utilization of the quantum battery.

\begin{figure*}[t]
\centering
\includegraphics[width=17cm,height=21.1cm,keepaspectratio]{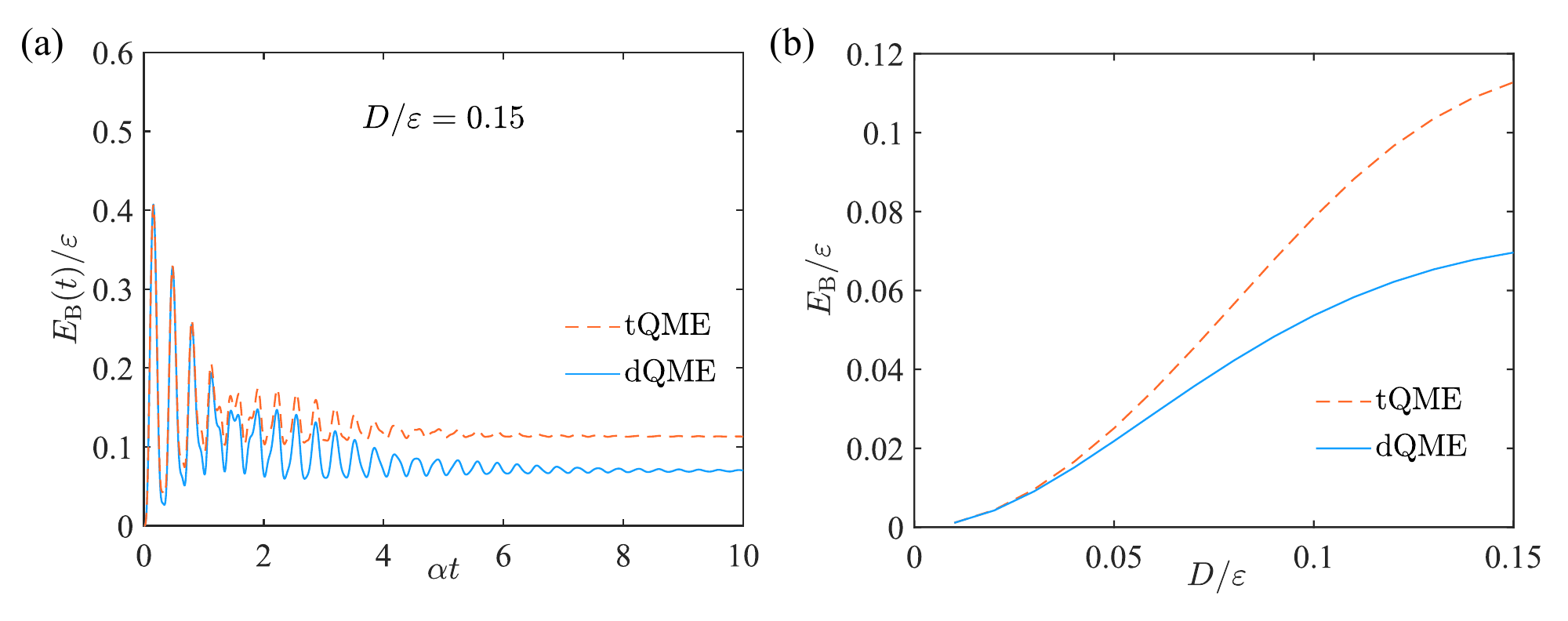}
\caption{ 
Comparisons of (a) the dynamics and
(b) approximate steady state (estimated at $\alpha{t}=20$) of the stored battery energy in the coupled-qubit model under dQME and tQME, 
with the initial state $|\psi(0){\rangle}=|\downarrow\downarrow{\rangle}$. In panel (a), $D=0.15\varepsilon$; panel (b) varies $D$.
Other system parameters are given by
$\varepsilon=1$, $\lambda=0.1\varepsilon$,
$\omega_d=0.8\varepsilon$,
$\alpha_l=\alpha_r=10^{-3}\varepsilon$, $\omega_c=10\varepsilon$, and
$k_BT_l=k_BT_r{=}10^{-5}\varepsilon$.
}\label{fig2}
\end{figure*}

\subsection{Driven coupled-qubit battery}

We consider the minimal bipartite quantum system, i.e.,
a coupled-qubit system, to investigate charging performance under the interplay between coherent driving and quantum dissipation.
The Hamiltonian of the driven coupled-qubit model is described as
\begin{align}\label{dQ2}
\hat{H}_{\textrm{Q2}}(t)&=\varepsilon\sum_{\mu=l,r}\hat{\sigma}_{\mu,+}\hat{\sigma}_{\mu,-}+\lambda(\hat{\sigma}_{l,+}\hat{\sigma}_{r,-}+\hat{\sigma}_{r,+}\hat{\sigma}_{l,-})\nonumber\\
&\quad-{D}(e^{-i\omega_dt}\hat{\sigma}_{l,+}+e^{i\omega_dt}\hat{\sigma}_{l,-}).
\end{align}
After the rotating transformation, the system Hamiltonian becomes
\begin{align}\label{Q2R}
    \hat{H}_{\rm Q2,\rm R}&=\Delta\sum_{\mu=l,r}\hat{\sigma}_{\mu,+}\hat{\sigma}_{\mu,-}+\lambda(\hat{\sigma}_{l,+}\hat{\sigma}_{r,-}+\hat{\sigma}_{r,+}\hat{\sigma}_{l,-})\nonumber\\
    &\quad-D(\hat{\sigma}_{l,+}+\hat{\sigma}_{l,-}),
\end{align} 
where $\hat{\sigma}_{\mu,\pm}=(\hat{\sigma}_{\mu,x}{\pm}i\hat{\sigma}_{\mu,y})/2$, with $\hat{\sigma}_{\mu,\alpha}~(\alpha=x,y,z)$ the Pauli operators, the detuning is $\Delta=\varepsilon-\omega_d$,
with $\varepsilon$ the Zeeman splitting energy of the qubits and $\omega_d$ the driving frequency,
$\lambda$ describes the interqubit coupling strength,
and $D$ is the driving amplitude.

For the coupled-qubit system, the energies of the battery and charger in the rotating frame can be straightforwardly specified as
\begin{subequations}
    \begin{align}
        E_{\rm B}(t)&=\varepsilon\textrm{Tr}_s\{\hat{\sigma}_{r,+}\hat{\sigma}_{r,-}\hat{\rho}_{s,\rm R}(t)\},~\label{Qubit:EBt}\\
       E_{\rm C}(t)&=\varepsilon\textrm{Tr}_s\{\hat{\sigma}_{l,+}\hat{\sigma}_{l,-}\hat{\rho}_{s,\rm R}(t)\}.~\label{Qubit:ECt}
    \end{align}
\end{subequations}
The battery energy can also be reexpressed as
\begin{align}
    E_{\rm B}/\varepsilon=P_{e}+(P_++P_-)/2-\rm{Re}(P_{+-}).
\end{align}
A general expression can also be obtained for the ergotropy of the battery qubit.
Writing the Hamiltonian and density operator of the battery as
$\hat{H}_{\rm B}=\varepsilon\hat{\sigma}_{r,+}\hat{\sigma}_{r,-}$
and
$\hat{\rho}_{\rm B}=(\mathrm{I}+\vec{S}{\cdot}{\hat{\sigma}})/2$,
with $\mathrm{I}$ the identity operator,
$\vec{S}=(S_x,S_y,S_z)$ the Bloch vector,
and
$\hat{\sigma}=(\hat{\sigma}_x,\hat{\sigma}_y,\hat{\sigma}_z)$
the vector of Pauli operators,
the ergotropy simplifies to
$\mathcal{E}_{\rm B}(t)=\frac{\varepsilon}{2}(|\vec{S}|+S_z)$,
which can be reexpressed as
\begin{align}\label{ego1}
\mathcal{E}_{\rm B}(t)=\frac{\varepsilon}{2}\Big(\sqrt{{\langle}\hat{\sigma}_{r,z}{\rangle}^2+4|{\langle}\hat{\sigma}_{r,+}{\rangle}|^2}+{\langle}\hat{\sigma}_{r,z}{\rangle}\Big).  
\end{align}

\begin{figure*}[t]
\centering
\includegraphics[width=17cm,height=21.1cm,keepaspectratio]{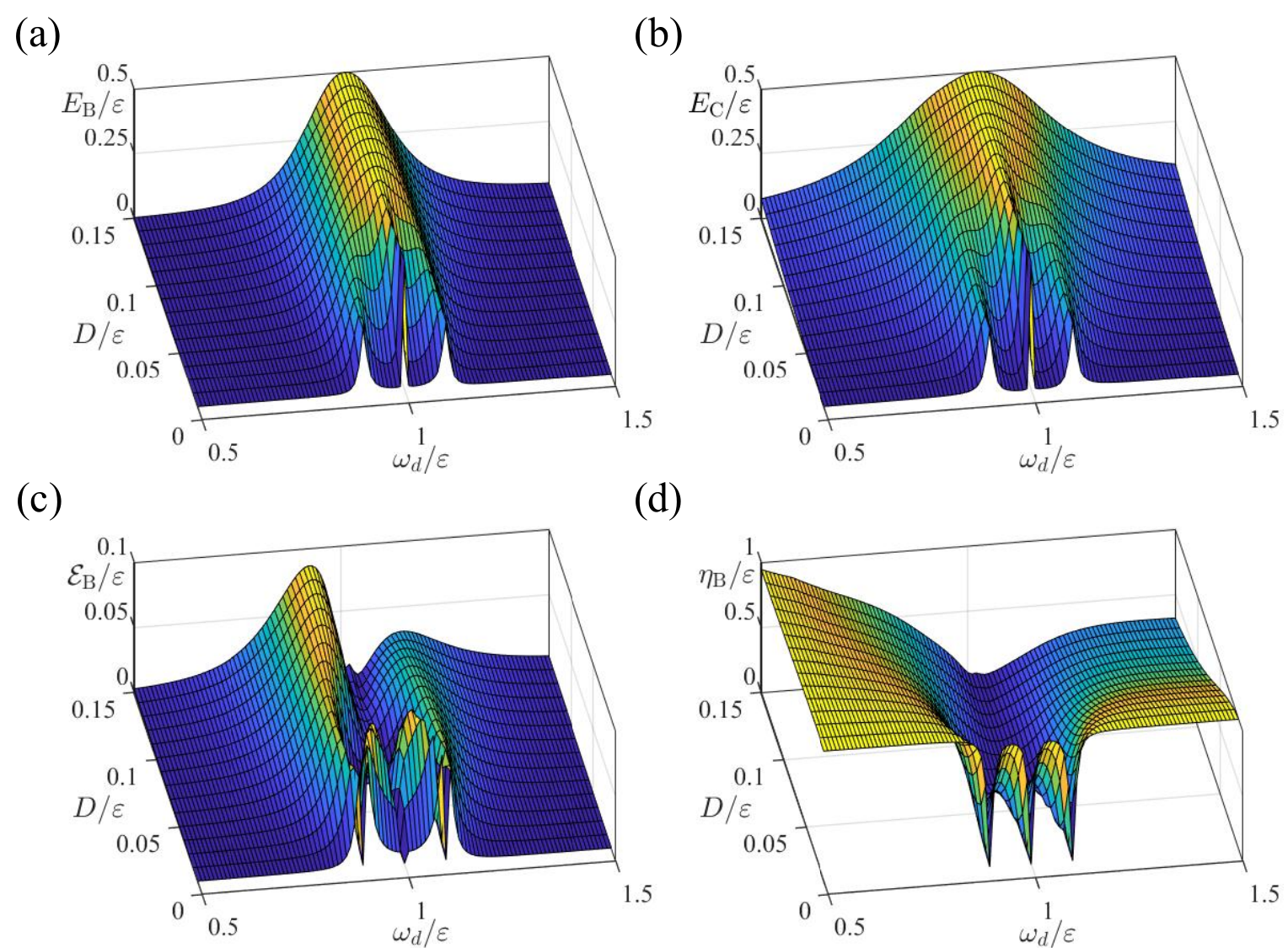}
\caption{ 
Influence of the driving amplitude $D$ and frequency $\omega_d$ on steady-state 
(a) battery energy $E_{\rm B}$,
(b) charger energy $E_{\rm C}$,
(c) ergotropy of the quantum battery $\mathcal{E}_{\rm B}$,
and (d) charging efficiency $\eta_{\rm B}=\mathcal{E}_{\rm B}/E_{\rm B}$.
Other system parameters are given by
$\varepsilon=1$,  $\lambda=0.1\varepsilon$,
$\alpha_l=\alpha_r=10^{-3}\varepsilon$, $\omega_c=10\varepsilon$,
and
$k_BT_l=k_BT_r=10^{-5}\varepsilon$.
}\label{fig3}
\end{figure*}

\begin{figure*}[t]
\centering
\includegraphics[width=17cm,height=21.1cm,keepaspectratio]{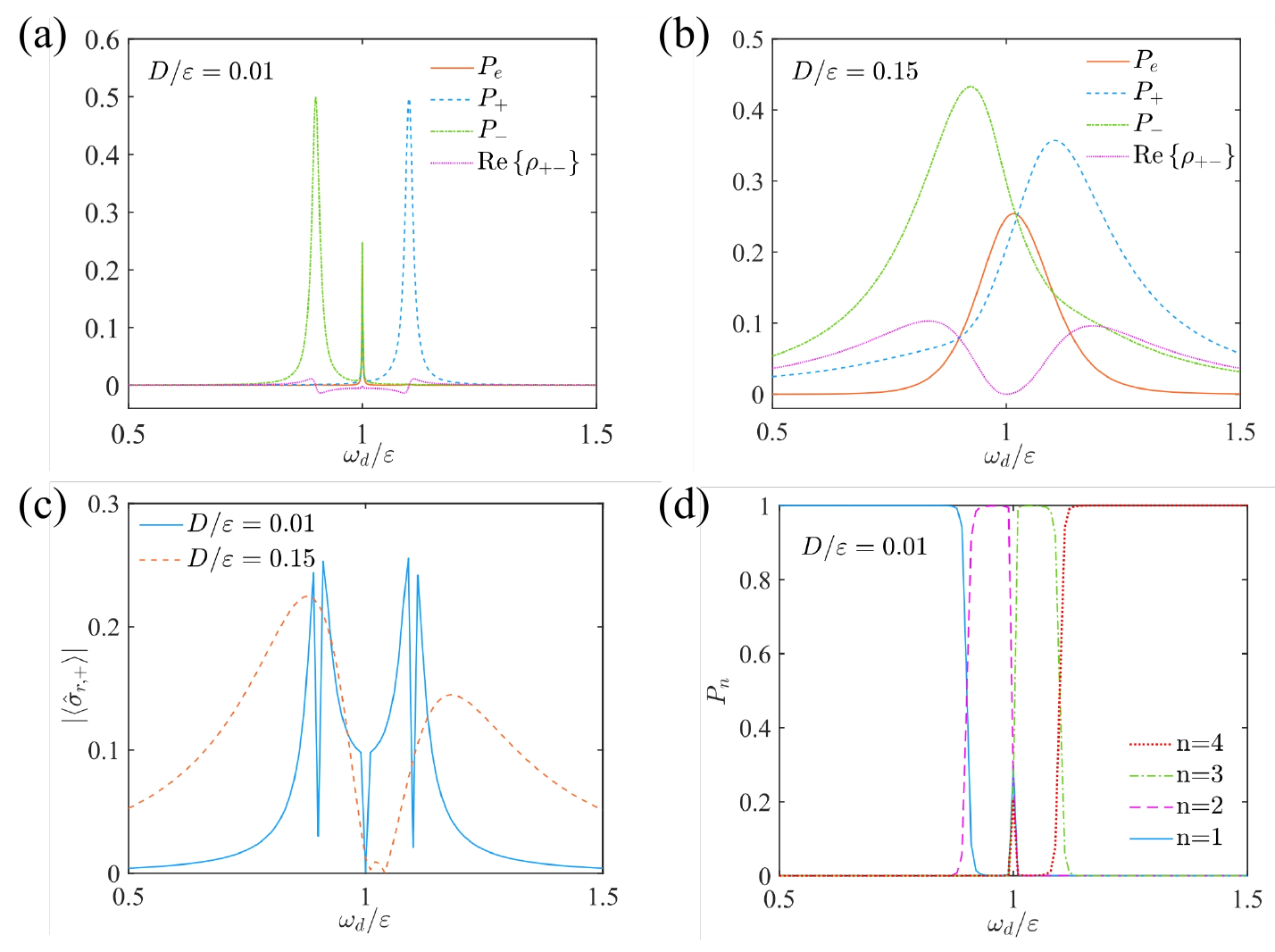}
\caption{ 
The influence of the driving frequency $\omega_d$ on the reduced density matrix elements
with (a) weak driving amplitude $D=0.01\varepsilon$ and
(b) strong driving amplitude $D=0.15\varepsilon$.
(c) Coefficients $|{\langle}\hat{\sigma}_{r,+}{\rangle}|$ and (d) steady-state populations $P_n={\langle}\phi_n|\hat{\rho}_{s,\rm R}|\phi_n{\rangle}~(n=1,2,3,4)$ as functions of $\omega_d$.
Other system parameters are the same as in Fig.~\ref{fig3}.
}\label{fig4}
\end{figure*}

\subsection{Charging performance}

We first investigate transient charging in the coupled-qubit model using dQME~(\ref{qme1}) and compare it with tQME~(\ref{dme0}) in Fig.~\ref{fig2}.
As expected, the charging energy $E_{\rm B}$~(\ref{Qubit:EBt}) shows dissipative transient dynamics and reaches a steady state under both methods.
However, Fig.~\ref{fig2}(a) shows that $E_{\rm B}$ obtained with dQME exhibits longer-lasting oscillations and stronger backflow from the battery to the charger than its tQME counterpart. 
After long-time evolution, $E_{\rm B}$ at a weak driving amplitude ($D/\varepsilon{\ll}1$) under dQME is expected to overlap with the one under tQME.
However, $E_{\rm B}$ is dramatically suppressed under dQME as $D$ increases (e.g., $D=0.1\varepsilon$), as shown in Fig.~\ref{fig2}(b).
This distinction clearly demonstrates the necessary correction of the dressed picture based on $\hat{H}_{s,\rm R}$ after the rotating transformation, in particular for the steady-state coherent energy transfer behaviors.

Next, we analyze the influence of the driving amplitude $D$ and frequency $\omega_d$ on steady-state charging performance in Fig.~\ref{fig3}.
The battery energy $E_{\rm B}$ with a weak driving amplitude (e.g., $D/\varepsilon=0.01$) shows three resonances of efficient energy transfer, as shown in Fig.~\ref{fig3}(a).
To exhibit the microscopic energy exchange processes, we define the Hilbert space $\mathcal{B}=\{|g{\rangle}=|\downarrow\downarrow{\rangle},|\pm{\rangle},|e{\rangle}=|\uparrow\uparrow{\rangle}\}$, with symmetric and antisymmetric qubit states 
$|\pm{\rangle}=(|\uparrow\downarrow{\rangle}{\pm}|\downarrow\uparrow{\rangle})/\sqrt{2}$.
Then a single-excitation resonance, i.e., $|g{\rangle}{\rightarrow}|s{\rangle}$, occurs when $|g{\rangle}$ becomes nearly
degenerate with one of the single-excitation states $|s{\rangle}$, with $s=\pm{1}$.
Including the weak driving terms, the effective eigenvalues are obtained as
$E_{s,\pm}=({\Delta + s\lambda
\pm \sqrt{(\Delta + s\lambda)^2 + 2D^2}})/{2}$,
and the corresponding eigenvectors are given by
$|\phi_{s,-}{\rangle}=\cos(\theta_s/2)|g{\rangle}+\sin(\theta_s/2)|s{\rangle}$
and
$|\phi_{s,+}{\rangle}=-\sin(\theta_s/2)|g{\rangle}+\cos(\theta_s/2)|s{\rangle}$,
with the angle $\theta_s=\arctan(\sqrt{2}D/(\Delta+s\lambda))$.
Accordingly, the single-excitation resonance relation can be obtained as
$\theta_{s}{\rightarrow}\pi/2$, i.e.,
$\omega_d{\approx}\varepsilon{+}s\lambda$, where $|g{\rangle}$ and 
$|s{\rangle}$ become highly entangled.
Since each single-excitation state has battery occupation $1/2$, its excitation provides a direct route for storing energy in the right qubit.
It is known that the battery energy in the $\mathcal{B}$ space can be reexpressed as
\begin{align}\label{Eb2}
    \frac{E_{\rm B}}{\varepsilon}
= P_e + \frac{P_+ + P_-}{2}
- \operatorname{Re}\{\rho_{+-}\},
\end{align}
with the population $P_a = {\langle}a|\hat{\rho}_{s,\rm{R}}|a{\rangle}$,
and the coherence
$\rho_{+-} = {\langle}+|\hat{\rho}_{s,\rm{R}}|-{\rangle}$.
The marked increase in $P_{\pm}$ in Fig.~\ref{fig4}(a) supports the single-excitation resonance mechanism.

The central resonance has a distinct origin.
$|g\rangle$ and $|e\rangle$ around $\Delta=0$ are nearly degenerate, whereas 
$|{s}\rangle$ is detuned by approximately $s{\lambda}$. 
However, there is no direct tunneling matrix element between $|g\rangle$ and $|e\rangle$.
Their coupling instead arises at second order in the drive through the two successive paths $|g\rangle{\rightarrow}|s\rangle{\rightarrow}|e\rangle$. 
More specifically, we express $\hat{H}_{\rm Q2,R}=\hat{H}_0+\hat{V}$,
with $\hat{H}_0=\Delta\sum_{\mu=l,r}\hat{\sigma}_{\mu,+}\hat{\sigma}_{\mu,-}
+\lambda\bigl(\hat{\sigma}_{l,+}\hat{\sigma}_{r,-}+\hat{\sigma}_{l,-}\hat{\sigma}_{r,+}\bigr)$
and
$\hat{V}=-D\bigl(\hat{\sigma}_{l,+}+\hat{\sigma}_{l,-}\bigr)$.
We then apply the Schrieffer-Wolff transformation 
$\hat{H}^\prime=e^{\hat{S}}(\hat{H}_0+\hat{V})e^{-\hat{S}}$
with the operator determined by $[\hat{S},\hat{H}_0]=-\hat{V}$.
The effective system Hamiltonian in second-order perturbation theory is given by
\begin{align}
\bigl(\hat{H}_{\mathrm{eff}}\bigr)_{ab}
&{\approx}E_a^{(0)}\delta_{ab}\nonumber\\
&+\frac12\sum_{s=\pm}V_{as}V_{sb}
\left[
\frac{1}{E_a^{(0)}-E_s^{(0)}}
+\frac{1}{E_b^{(0)}-E_s^{(0)}}
\right],
\end{align}
where the indices $a,b\in\{g,e\}$ and $E^{(0)}_a$ is the eigenvalue of $\hat{H}_0$.
Thus, around $\Delta{=}0$ it is straightforward to find the simplified expression
\begin{align}\label{doubE}
\hat{H}_{\mathrm{eff}}
{\approx}-\frac{D^2}{\lambda}
\bigl(\ket{g}\bra{e}+\ket{e}\bra{g}\bigr).    
\end{align}
This demonstrates the effective double-excitation tunneling $|g{\rangle}{\dashrightarrow}|e{\rangle}$.
The significant excitation of $P_e$ in Fig.~\ref{fig4}(a) also supports this double-excitation resonance mechanism.
We note that the central correction of $|\pm\rangle$ mediating the effective coupling in Eq.~(\ref{doubE}) requires $D\ll\lambda$ near resonance, which is also the limitation of the Schrieffer-Wolff approximation.

As the driving amplitude becomes comparable to the interqubit coupling,
e.g., $D\sim\lambda$, the side peaks near $\omega_d\simeq\varepsilon\pm\lambda$
gradually become less distinguishable.
The steady-state battery energy develops
a broad central peak around $\omega_d\simeq\varepsilon$, as shown in
Fig.~\ref{fig3}(a). In this regime, the drive-induced couplings are
comparable to the relevant unperturbed energy separations, leading to
apparent mixing within the full four-state manifold
$\mathcal{B}=\{|g\rangle,|+\rangle,|-\rangle,|e\rangle\}$.
Consequently, the weak-driving description of separate
single-excitation resonances and the effective double-excitation resonance may break down, as shown in Fig.~\ref{fig4}(b).
Although the driving field acts
locally on the left qubit, energy transfer to the right qubit 
should involve intermediate states to generate interqubit exchange coupling and the resulting mixing within
the $\mathcal{B}$ space, i.e., the emergence of quantum coherence $\rm{Re}\{\rho_{+-}\}$. 
The broad central feature therefore reflects the response
of the full driven four-state system. 
Its shape is determined jointly
by coherent mixing and the drive-dependent dissipation in Eq.~(\ref{qme1}),
which establishes steady-state populations and quantum coherence entering
the battery energy in Eq.~(\ref{Eb2}). 
For the parameters considered, their
combined effect naturally produces the crossover from three resolved peaks
to a single broad peak. 
Meanwhile, the evolution of the charger energy
$E_{\rm C}$ in Fig.~\ref{fig3}(b) reflects the analogous underlying four-state steady-state dynamics.

Figure~\ref{fig3}(c) shows the ergotropy $\mathcal{E}_{\rm B}$ at steady state by tuning the driving frequency and amplitude.
With a weak driving amplitude, $\mathcal{E}_{\rm B}$ shows sudden suppression at the three resonance positions, indicating inefficient charging.
Specifically, at $\omega_d{\approx}\varepsilon$, the two-photon-resonance
transition dramatically enhances the excitation of the right qubit, leading to ${\langle}\hat{\sigma}_{r,z}{\rangle}{\approx}0$ (i.e., $E_{\rm B}/\varepsilon=({\langle}\hat{\sigma}_{r,z}{\rangle}+1)/2{\approx}0.5$). 
Meanwhile, 
\begin{align}
|{\langle}\hat{\sigma}_{r,+}{\rangle}|{\approx}(D/\sqrt{\lambda^2+4D^2})|(P_2+P_3-P_1-P_4)|    
\end{align}
disappears (see Fig.~\ref{fig4}(c) and Appendix~\ref{Append:A}), with 
$P_n={\langle}\phi_n|\hat{\rho}_{s,\rm R}|\phi_n{\rangle}{\approx}1/4$ as shown in Fig.~\ref{fig4}(d).
Based on the expression of $\mathcal{E}_{\rm B}$ at Eq.~(\ref{ego1}), 
the ergotropy is dramatically suppressed.
Around the two side valleys  $\omega_d{\approx}\varepsilon{\mp}\lambda$,
the energy exchange processes in the low-temperature limit are mainly dominated by two near-resonant states $|\downarrow\downarrow{\rangle}$
and $|s{\rangle}~(s=\pm)$.
The steady state is approximately expressed as
$\hat{\rho}_{s,\rm R}{\approx}(|\downarrow\downarrow{\rangle}{\langle}\downarrow\downarrow|+|s{\rangle}{\langle}s|)/2$,
as also shown in Fig.~\ref{fig4}(a).
This leads to the reduced density operator of the battery component
$\hat{\rho}_{\rm B}=3|\downarrow{\rangle}{\langle}\downarrow|/4+|\uparrow{\rangle}{\langle}\uparrow|/4$.
Hence, ${\langle}\hat{\sigma}_{r,+}{\rangle}$ vanishes.
Meanwhile, ${\langle}\hat{\sigma}_{r,z}{\rangle}{\approx}-1/2$.
Therefore, the ergotropy is expected to vanish sharply.
These three sudden drops of $\mathcal{E}_{\rm B}$ yield
the corresponding low charging efficiency $\eta_{\rm B}$, as shown in Fig.~\ref{fig3}(d).
As the driving amplitude increases, e.g., $D=0.15\varepsilon$, the absence of ${\langle}\hat{\sigma}_{r,+}{\rangle}$ persists around $\omega_d{\approx}\varepsilon$, which directly leads to 
a valley in $\mathcal{E}_{\rm B}$.
However, near the side resonance $\omega_d{\approx}\varepsilon{-}\lambda$, the element
 $|{\langle}\hat{\sigma}_{r,+}{\rangle}|$ becomes finite, which strongly increases the ergotropy.
Considering the condition ${\langle}\hat{\sigma}_{r,z}{\rangle}{\approx}-1$ (i.e., $E_{\rm B}/\varepsilon{\ll}1$), the ergotropy at Eq.~(\ref{ego1}) could be simplified as
\begin{align}
    \mathcal{E}_{\rm B}{\approx}\varepsilon|{\langle}\hat{\sigma}_{r,+}{\rangle}|^2.
\end{align}
Thus, the peak around $\omega_d{\approx}\varepsilon-\lambda$ is higher than that around $\omega_d{\approx}\varepsilon+\lambda$, mainly due to the asymmetric behavior of ${\langle}\hat{\sigma}_{r,+}{\rangle}$. 
Consequently, the charging efficiency at a large driving amplitude also exhibits a pronounced asymmetry by tuning $\omega_d$.

\begin{figure*}[t]
\centering
\includegraphics[width=17cm,height=21.1cm,keepaspectratio]{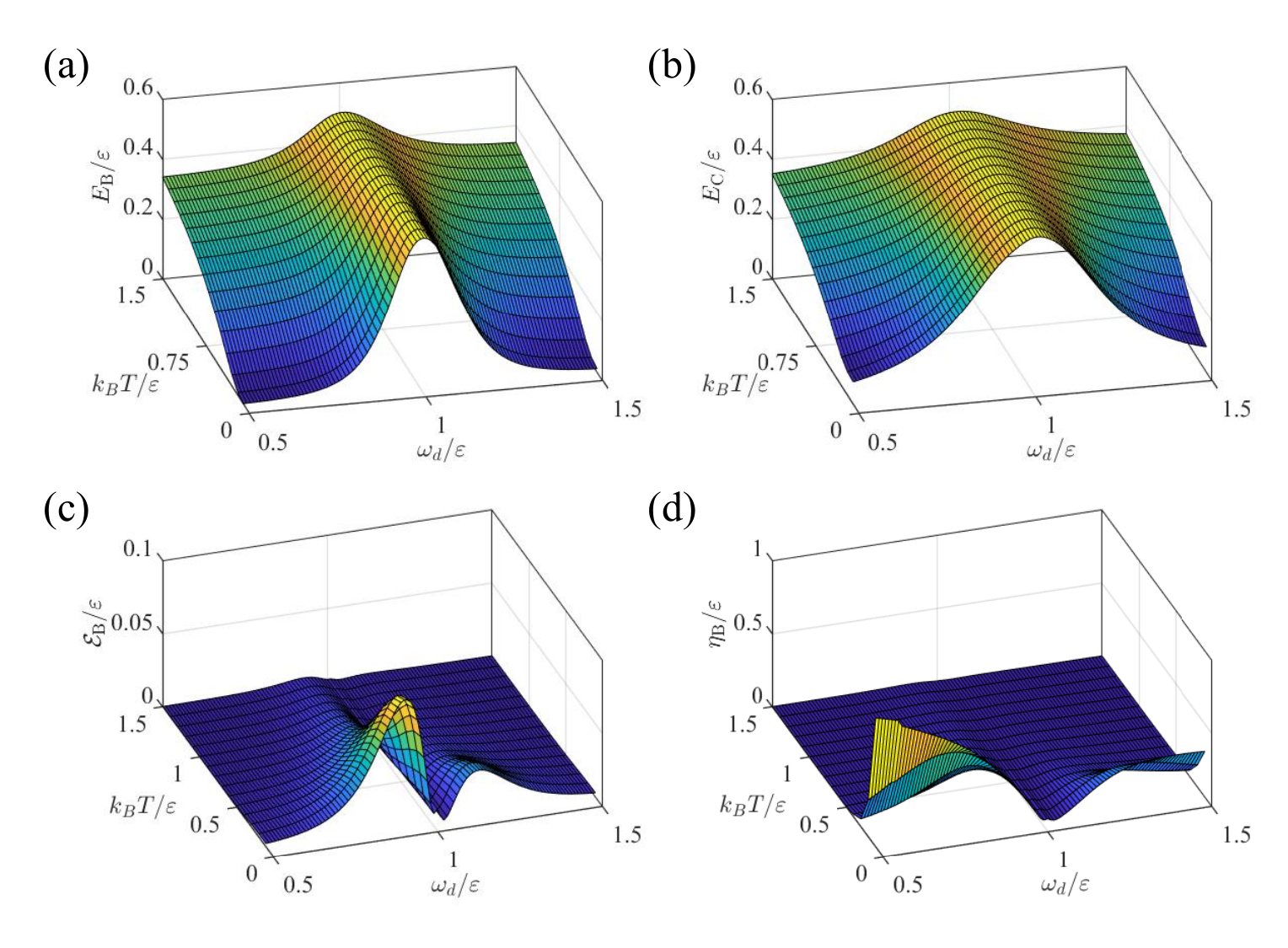}
\caption{ 
Influence of the reservoir temperatures $T_l=T_r=T$
and driving frequency $\omega_d$ on the battery energy $E_{\rm B}$, charger energy  $E_{\rm C}$, 
the ergotropy of the quantum battery $\mathcal{E}_{\rm B}$, 
and charging efficiency $\eta_{\rm B}$
at steady state.
Other system parameters are given by
$\varepsilon=1$,  $\lambda=0.1\varepsilon$,
$D=0.15\varepsilon$,
$\alpha_l=\alpha_r=10^{-3}\varepsilon$, 
and $\omega_c=10\varepsilon$.
}\label{fig5}
\end{figure*}

We also investigate the thermal effect on quantum charging performance at steady state in the driven-dissipative coupled-qubit system in Fig.~\ref{fig5}.
As the temperature increases,
the charging energy in Fig.~\ref{fig5}(a) is generally enhanced,
and the peak of $E_{\rm B}$ around $\omega_d{\approx}\varepsilon$ becomes robust.
The temperature dependence of the charger energy $E_{\rm C}$ in Fig.~\ref{fig5}(b) is quite similar.
The ergotropy $\mathcal{E}_{\rm B}$ in Fig.~\ref{fig5}(c) is monotonically suppressed by increasing $k_BT$, which demonstrates that thermal effects are detrimental to the extractable work of the quantum battery.
Consequently, the charging efficiency is also dramatically suppressed by increasing temperature, as shown in Fig.~\ref{fig5}(d).

\section{Driven incoherent energy transfer}\label{sec:result2}

\subsection{Expression of steady-state energy current}

We use the dQME in Eq.~(\ref{qme1}) to analyze quantum transfer in driven open quantum systems.
By defining the energy current 
\begin{align}\label{mc0}
J_\mu(t){\equiv}\frac{d}{dt}\langle \hat{H}_b^\mu\rangle
=i\left\langle[\hat{V}_\mu(t),\hat{H}_b^\mu]\right\rangle
\end{align}
with $\hat{V}_\mu(t)$ in Eq.~(\ref{vR}),
we obtain the microscopic current operator as
$\hat{\mathcal{I}}_\mu(t)
=i\sum_k\omega_{k\mu}\left[
g_{k\mu}^*e^{i\omega_dt}\hat{b}_{k\mu}\hat{A}_\mu^\dagger
-g_{k\mu}e^{-i\omega_dt}\hat{b}_{k\mu}^\dagger \hat{A}_\mu
\right]$.
Based on perturbation theory (see Appendix~\ref{Append:B}),
a concise expression for the energy current is
\begin{equation}
J_\mu(t)=2\mathrm{Re}\left\{\textrm{Tr}\left[
\left(\hat{A}_\mu^\dagger \hat{\mathcal K}_{\mu,-}
-\hat{A}_\mu \hat{\mathcal K}_{\mu,+}^{\dagger}\right)\hat{\rho}_{s,\rm R}(t)
\right]\right\},\label{current1}
\end{equation}
where the dressed operator is
\begin{align}
    \hat{\mathcal K}_{\mu,\pm}
=\frac12\sum_{m,n}
q_{mn}\gamma_\mu(q_{mn})n_{\mu,\pm}(q_{mn})
A^\mu_{mn}|\phi_m\rangle\langle\phi_n|,
\end{align}
with $n_{\mu,+}(\omega)=n_\mu(\omega)$ and $n_{\mu,-}(\omega)=1+n_\mu(\omega)$.
The exchanged energy is $q_{mn}=\omega_d+E_n-E_m$,
and the transition coefficient is 
$A^\mu_{mn}={\langle}\phi_m|\hat{A}_\mu|\phi_n{\rangle}$.
At steady state, energy conservation gives the current into the driving terminal as
\begin{align}
    J_p=-(J_l+J_r).
\end{align}

In contrast, based on the tQME at Eq.~(\ref{dme0}), the eigenstates are obtained from the undriven component of the quantum systems.
Therefore, based on the incoherent energy exchange processes,
the steady-state expression of the energy current into the $\mu$-th reservoir is given by
\begin{align}
       J^\prime_\mu&=-\sum_{m,m^\prime}\bigl\{\Gamma^\mu_{-}(\Delta_{m^{\prime}m}){\rm{Tr}}\{\hat{H}_s\mathcal{L}[|\psi_m{\rangle}{\langle}\psi_{m^\prime}|]\hat{\rho}_{s}\}\nonumber\\
       &\qquad+\Gamma^\mu_{+}(\Delta_{m^{\prime}m}){\rm{Tr}}\{\hat{H}_s\mathcal{L}[|\psi_{m^\prime}{\rangle}{\langle}\psi_{m}|]\hat{\rho}_{s}\}\} 
       ,~\label{Jmup}
\end{align}
with $\hat{H}_s$ the static component of the driven quantum system
and $\hat{\rho}_{s}$ the reduced system density operator in the original Schr\"odinger picture.
One would find the straightforward expression of the driving-induced energy current $J^\prime_p=-(J^\prime_l+J^\prime_r)$.

\subsection{Driven quantum energy transfer}

\subsubsection{Driven single qubit}

\begin{figure*}[t]
\centering
\includegraphics[width=17cm,height=21.1cm,keepaspectratio]{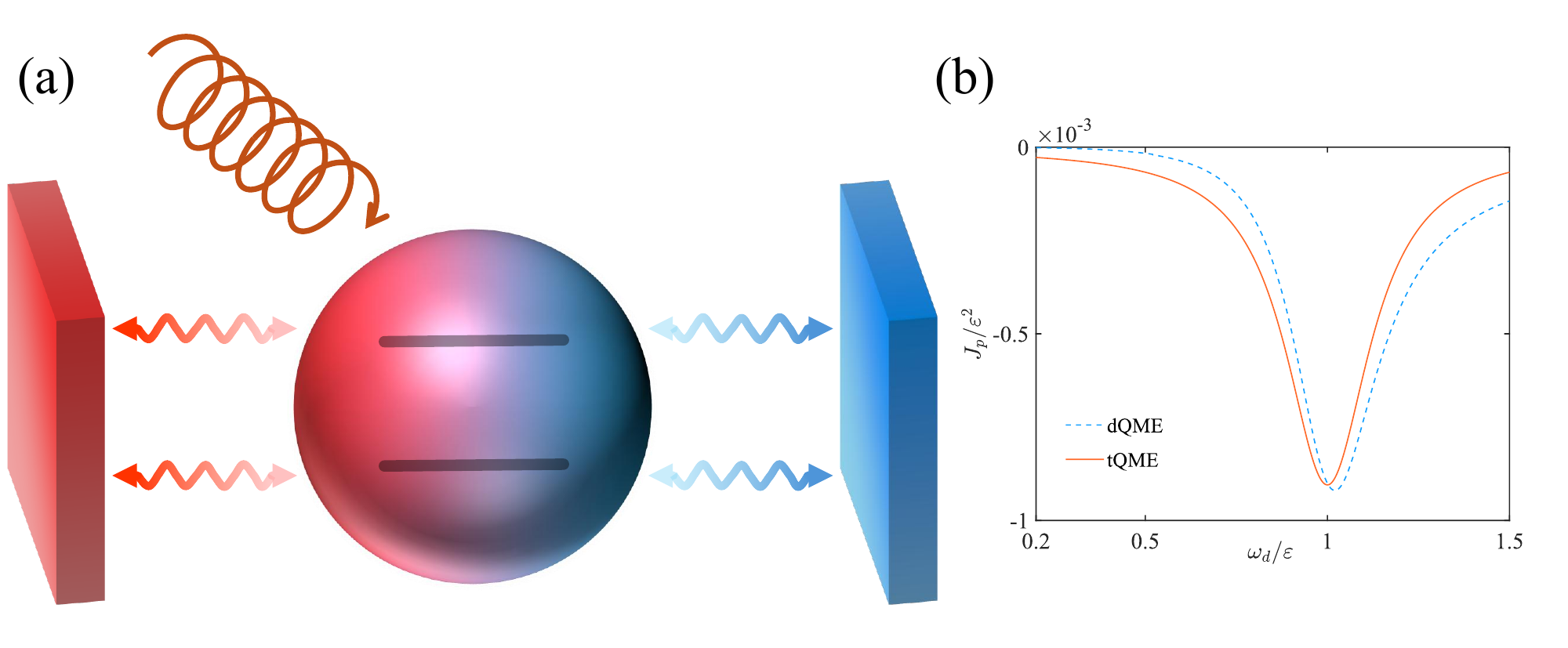}
\caption{
(a) Schematic of a nonequilibrium driven spin-boson model,
where the central sphere denotes a driven two-level qubit,
the brown spiral line with an arrow represents the external driving field applied to the quantum system,
and the red pane on the left and blue pane on the right represent the
bosonic thermal reservoirs.
(b) Comparison of the energy current into the driving terminal, $J_p$, with dQME and tQME, by tuning the driving frequency $\omega_d$.
Other system parameters are given by
$\varepsilon=1$, $D=0.1\varepsilon$,
$\alpha_l=\alpha_r=10^{-3}\varepsilon$, $\omega_c=10\varepsilon$,
$k_BT_l=1.2\varepsilon$, and $k_BT_r=0.4\varepsilon$.
}\label{fig6}
\end{figure*}

We first analyze energy transfer of the nonequilibrium driven spin-boson model~\cite{carrega2016prl,wang2017pra,wang2022fop}, where a two-level qubit with coherent driving interacts with two individual bosonic thermal reservoirs, as shown in Fig.~\ref{fig6}(a).
The Hamiltonian reads
$\hat{H}_{\rm dSBM}(t)={\varepsilon}\hat{\sigma}_+\hat{\sigma}_--D(e^{-i\omega_d{t}}\hat{\sigma}_++H.c.)+
\sum_{\mu=l,r;k}[\omega_{k,\mu}\hat{b}^\dag_{k,\mu}\hat{b}_{k,\mu}+(g_{k,\mu}\hat{b}^\dag_{k,\mu}\hat{\sigma}_-+H.c.)]$,
where the Pauli operators under the spin basis $\{|\uparrow{\rangle}, |\downarrow{\rangle}\}$ denote $\hat{\sigma}_+=|\uparrow{\rangle}{\langle}\downarrow|$
and $\hat{\sigma}_-=(\hat{\sigma}_+)^\dag$.
After the rotation, the driven spin-boson Hamiltonian becomes
\begin{align}\label{HdQ}
    \hat{H}_{\rm dQ1}=\Delta\hat{\sigma}_+\hat{\sigma}_--D(\hat{\sigma}_++\hat{\sigma}_-),
\end{align}
where the detuning is $\Delta=\varepsilon-\omega_d$.

We then compare the energy current into the driving terminal, $J_p$, under dQME with the one under tQME in Fig.~\ref{fig6}(b).
It is found that the currents, e.g., $J_p$, with these two methods show a clear difference.
These results indicate that retaining the driving phase in system-reservoir interactions is indispensable to correctly characterize microscopic energy exchange processes.
Therefore, dQME is generally suitable for investigating driven quantum energy transfer.

\begin{figure*}[t]
\centering
\includegraphics[width=17cm,height=21.1cm,keepaspectratio]{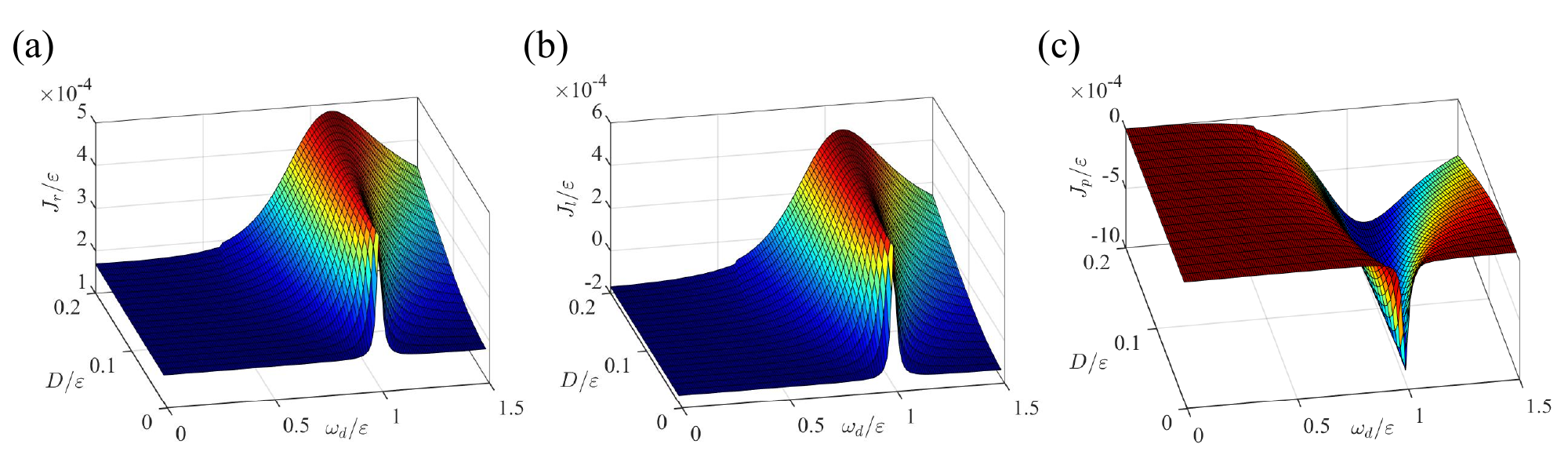}
\caption{
Influence of the driving amplitude and frequency on energy currents (a) $J_r$, (b) $J_l$, and (c) $J_p$, in the driven spin-boson model.
Other system parameters are given by
$\varepsilon=1$,
$\alpha_l=\alpha_r=10^{-3}\varepsilon$, $\omega_c=10\varepsilon$,
$k_BT_l=1.2\varepsilon$, and $k_BT_r=0.4\varepsilon$.
}\label{fig7}
\end{figure*}

Next, we show the influence of driving frequency $\omega_d$ and amplitude $D$ on energy currents in Fig.~\ref{fig7}.
Analytical expressions for energy currents into the thermal reservoirs (see Appendix~\ref{Append:C} for the details) can be approximately described as
\begin{align}\label{jmuq1}
    J_\mu&=(\omega_d+\Lambda)[\Gamma^\mu_-(\omega_d+\Lambda)P_+-\Gamma^\mu_+(\omega_d+\Lambda)P_-]\nonumber\\
    &+(\omega_d-\Lambda)[\Gamma^\mu_-(\omega_d-\Lambda)P_--\Gamma^\mu_+(\omega_d-\Lambda)P_+]\nonumber\\
    &+\omega_d[\Gamma^\mu_-(\omega_d)-\Gamma^\mu_+(\omega_d)],
\end{align}
where the steady-state populations become
$P_+={\Gamma_+}/{(\Gamma_++\Gamma_-)}$
and
$P_-={\Gamma_-}/{(\Gamma_++\Gamma_-)}$,
collective rates denote $\Gamma_+=\sum_{\mu=l,r;q={\pm}1}\Gamma^\mu_{q}(\omega_d+q\Lambda)$
and
$\Gamma_-=\sum_{\mu=l,r;q={\pm}1}\Gamma^\mu_{q}(\omega_d-q\Lambda)$,
and the rates are
$\Gamma^\mu_{\pm}(\omega_d+\Lambda)=\pm\gamma_\mu(\omega_d+\Lambda)n_\mu(\pm(\omega_d+\Lambda))\cos^4\frac{\theta}{2}$,
$\Gamma^\mu_{\pm}(\omega_d-\Lambda)=\pm\gamma_\mu(\omega_d-\Lambda)n_\mu(\pm(\omega_d-\Lambda))\sin^4\frac{\theta}{2}$,
and
$\Gamma^\mu_\pm(\omega_d)=\pm\gamma_\mu(\omega_d)n_\mu(\pm\omega_d)(\sin{\theta}/2)^2$,
with energy gap $\Lambda=\sqrt{\Delta^2+4D^2}$ and 
the angle $\tan\theta=2D/\Delta$.
Energy currents and microscopic transition rates clearly exhibit three energy exchange channels with resonant energies $(\omega_d{\pm}\Lambda)$ and $\omega_d$.
At weak driving amplitude, e.g., $D=0.01\varepsilon$,
the heat currents $J_\mu~(\mu=r,l,p)$ are optimally enhanced as $\omega_d{\approx}\varepsilon$,
whereas they are dramatically weakened once the driving frequency moves away from $\varepsilon$.
When the driving amplitude is tuned up, the peaks of heat currents are gradually broadened, in particular at $\omega_d>\varepsilon$.
This demonstrates the pumping effect on the energy currents.
Therefore, we conclude that comparatively strong driving enhances quantum energy transfer.

\subsubsection{Driven coupled qubits}

\begin{figure*}[t]
\centering
\includegraphics[width=17cm,height=21.1cm,keepaspectratio]{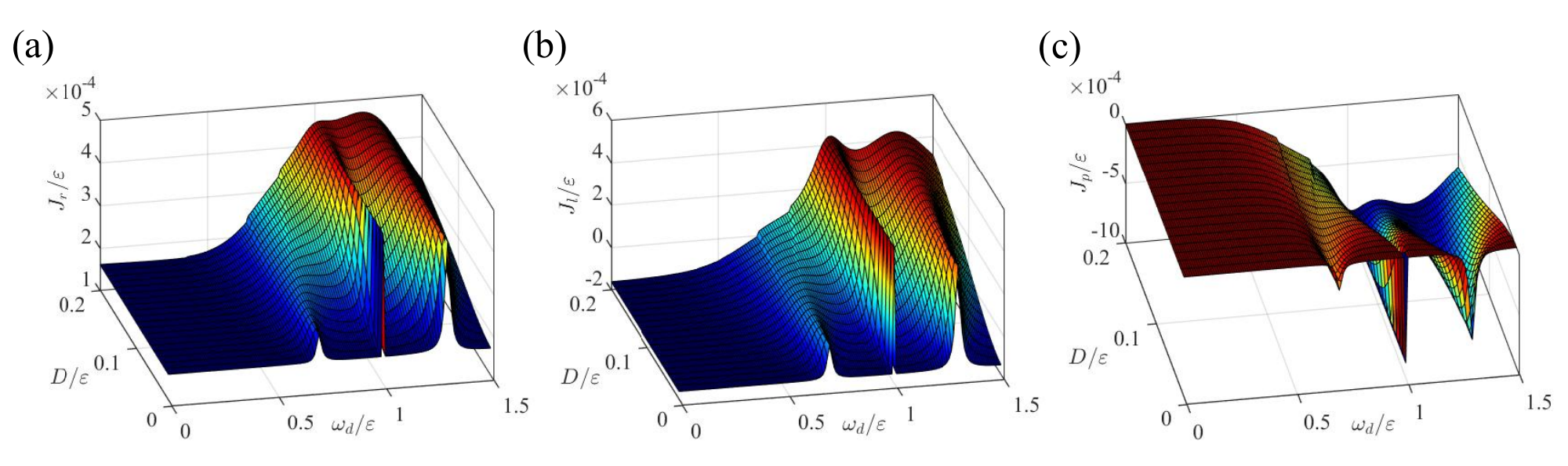}
\caption{
Influence of the driving amplitude and frequency on energy currents (a) $J_r$, (b) $J_l$, and (c) $J_p$, in the driven coupled-qubit system.
Other system parameters are given by
$\varepsilon=1$,
$\lambda=0.3\varepsilon$,
$\alpha_l=\alpha_r=10^{-3}\varepsilon$, $\omega_c=10\varepsilon$,
$k_BT_l=1.2\varepsilon$, and $k_BT_r=0.4\varepsilon$.
}\label{fig8}
\end{figure*}

We also consider energy transfer in two coupled qubits
to show the applicability of dQME to coupled quantum systems, where the driving field is applied to the left qubit.
After the rotating transformation, the driven system Hamiltonian is described as
\begin{align}
    \hat{H}_{\rm dQ2}&=\Delta\sum_{\mu=l,r}\hat{\sigma}_{\mu,+}\hat{\sigma}_{\mu,-}+\lambda(\hat{\sigma}_{l,+}\hat{\sigma}_{r,-}+\hat{\sigma}_{r,+}\hat{\sigma}_{l,-})\nonumber\\
    &-D(\hat{\sigma}_{l,+}+\hat{\sigma}_{l,-}).
\end{align}    

Based on dQME, the influence of the driving frequency and amplitude on
all energy currents, i.e., $J_r$, $J_l$, and $J_p$, is analyzed
in Fig.~\ref{fig8}.
At a weak driving amplitude (e.g., $D/\varepsilon=0.01$), three enhanced peaks are clearly observed in all currents at resonant positions $\omega_d{\approx}\varepsilon$
and $\omega_d{\approx}\varepsilon{\pm}\lambda$,
which indicate efficient energy pumping processes in the coupled-qubit system.
By increasing the driving amplitude, e.g., $D/\varepsilon=0.1$, these enhancements of energy currents induced by resonant driving persist and broaden, in particular around $\omega_d{\approx}\varepsilon+\lambda$.
Therefore, increasing the driving amplitude is beneficial for optimizing driving-induced energy flows.

\section{Conclusion}\label{sec:conclusion}

In this work, we developed a driven quantum master equation (dQME) to study coherent and dissipative energy transfer in driven open quantum systems. The approach adopts the eigenstates of the full driven Hamiltonian in the rotating frame and retains the driving phase in the system-reservoir interactions.  
Therefore, this approach includes the effect of coherent driving in both the coherent evolution and dissipative exchange processes. 
By comparing with the traditional tQME, dQME shows clear distinctions in charging dynamics, steady-state energy storage, and driven energy currents.

For coherent energy charging performance, we studied a driven coupled-qubit battery. Under weak driving, the stored energy shows three resonance peaks near $\omega_d=\varepsilon\pm\lambda$ associated with single-excitation transitions and  $\omega_d=\varepsilon$  associated with a double-excitation transition. 
The underlying mechanisms are demonstrated as highly entangled states between $|g{\rangle}$ and $|\pm{\rangle}$,
and effective double-excitation tunneling, respectively.
Strong driving merges the three peaks into a broad peak near the central resonance, mainly due to state mixing within the $\mathcal{B}$ space. 
We note that enhanced energy storage does not always lead to more extractable work. 
In the weak-driving limit, the ergotropy and charging efficiency drop at the three resonances as the battery coherence is suppressed. 
The central dip persists with stronger driving, while finite coherence supports extractable work away from the resonant position. 
Increasing the reservoir temperature raises the stored energy, but reduces the ergotropy and charging efficiency.

For dissipative energy transfer, we studied a driven single qubit and two coupled qubits connected to thermal reservoirs. The analytical expressions for energy currents in the single-qubit model reveal three energy exchange channels associated with the driving frequency and the dressed-state energy gap. 
Resonant driving enhances energy pumping.
Moreover, the coupling between two qubits produces additional side resonances of energy currents. 
Increasing the driving amplitude will broaden the frequency range over which the currents are generally enhanced. 
These results show why the driving contribution should be retained when describing incoherent energy exchange with the thermal reservoirs.

Within the Born-Markov approximation and the single-frequency driving considered here, the dQME provides a common framework for coherent energy charging and dissipative energy transfer. 
It shows how coherent driving and quantum dissipation cooperatively control stored energy, extractable work, and energy currents in driven open quantum systems.

\begin{acknowledgments}

This work is supported by the Zhejiang Provincial Natural Science Foundation of China under Grant No. LZ25A050001 and
the National Natural Science Foundation of China (NSFC) under Grant No. 12674313.
Liwei Duan acknowledges support from the National Natural Science Foundation of China (NSFC) under Grant No. 12305032 and Zhejiang Provincial Natural Science Foundation of China under Grant No. LQ23A050003.

\end{acknowledgments}

\section*{Author Declarations}
\subsection*{Conflict of Interest}
The authors have no conflicts to disclose.
\subsection*{Author Contributions}
\textbf{Junran Kong:} Formal analysis (equal); Investigation (equal); Methodology (equal); Validation (equal); Writing -- original draft (equal).
\textbf{Yitian Chen:} Formal analysis (equal); Investigation (equal); Methodology (equal); Validation (equal).
\textbf{Huan Liu:} Writing -- review \& editing.
\textbf{Liwei Duan:} Conceptualization (equal); Investigation (equal); Validation (equal); Writing -- review \& editing (equal).
\textbf{Chen Wang:} Conceptualization (equal); Investigation (equal); Project administration; Validation (equal); Writing -- original draft (equal); Writing -- review \& editing (equal).

\section*{Data Availability}
The data that support the findings of this study are available from the corresponding author upon reasonable request.

\appendix

\section{Battery coherence at resonance}
~\label{Append:A}

For $\omega_d=\varepsilon$, i.e., $\Delta=0$, after the rotating transformation, the rotated system Hamiltonian becomes
\begin{align}
\hat{H}_{R}=\lambda(\hat{\sigma}_{l,+}\hat{\sigma}_{r,-}+\hat{\sigma}_{r,+}\hat{\sigma}_{l,-})-D\hat{\sigma}_{l,x}.
\end{align}
We define $|G{\rangle}=|\downarrow\downarrow{\rangle}$,
$|E{\rangle}=|\uparrow\uparrow{\rangle}$,
and $|\pm{\rangle}=(|\uparrow\downarrow{\rangle}{\pm}|\downarrow\uparrow{\rangle})/\sqrt{2}$.
Then, we further introduce the basis
$|u_0{\rangle}=(|G{\rangle}+|E{\rangle})/\sqrt{2}$,
$|v_0{\rangle}=(|G{\rangle}-|E{\rangle})/\sqrt{2}$,
$|u_1{\rangle}=|+{\rangle}$,
and
$|v_1{\rangle}=|-{\rangle}$.
Consequently, the rotated Hamiltonian is naturally decomposed into two blocks, 
$\hat{H}_u=[0,-D;-D,\lambda]$
and
$\hat{H}_v=[0,-D;-D,-\lambda]$.
Then, the eigenvalues of $\hat{H}_R$ are obtained as
$E_1=-(\lambda+\Lambda)/2$,
$E_2=(\lambda-\Lambda)/2$,
$E_3=(\Lambda-\lambda)/2$,
$E_4=(\lambda+\Lambda)/2$,
with $\Lambda=\sqrt{\lambda^2+4D^2}$.
The corresponding eigenvectors are given by
$|\phi_1{\rangle}=\sin(\theta/2)|v_0{\rangle}+\cos(\theta/2)|v_1{\rangle}$,
$|\phi_2{\rangle}=\cos(\theta/2)|u_0{\rangle}+\sin(\theta/2)|u_1{\rangle}$,
$|\phi_3{\rangle}=\cos(\theta/2)|v_0{\rangle}-\sin(\theta/2)|v_1{\rangle}$,
and
$|\phi_4{\rangle}=-\sin(\theta/2)|u_0{\rangle}+\cos(\theta/2)|u_1{\rangle}$, with the angle $\theta=\arctan(2D/\lambda)$.
The coefficients are given by
${\langle}\phi_1|\hat{\sigma}_{r,+}|\phi_1{\rangle}={\langle}\phi_4|\hat{\sigma}_{r,+}|\phi_4{\rangle}=-D/\Lambda$,
and
${\langle}\phi_2|\hat{\sigma}_{r,+}|\phi_2{\rangle}={\langle}\phi_3|\hat{\sigma}_{r,+}|\phi_3{\rangle}=D/\Lambda$.
Moreover, the steady-state coherences in the eigenbasis vanish in the absence of bath-induced coherence.
The density operator is approximately expressed as
$\hat{\rho}_{ss}\approx\sum_nP_n|\phi_n{\rangle}{\langle}\phi_n|$.
Hence
\begin{align}
    |{\langle}\hat{\sigma}_{r,+}{\rangle}|&{\approx}|\sum_nP_n{\langle}\phi_n|\hat{\sigma}_{r,+}|\phi_n{\rangle}|\nonumber\\
&{\approx}\frac{D}{\Lambda}|(P_2+P_3-P_1-P_4)|.
\end{align}
The populations are nearly equally distributed, i.e., $P_n{\approx}1/4$, due to the approximate wide-band condition $\gamma_\mu(\omega_d{\pm}\Lambda){\approx}\gamma_\mu(\omega_d)$.
Finally, we find that 
${\langle}\hat{\sigma}_{r,+}{\rangle}$ vanishes.

\section{Expression of steady-state energy flow}\label{Append:B}

Starting from the microscopic system-reservoir energy exchange processes,
we derive the energy current
flowing into the $\mu$th reservoir. 
Specifically, the Hamiltonian of the $\mu$th reservoir is
$\hat{H}_b^\mu=\sum_k\omega_{k\mu}\hat{b}_{k\mu}^\dagger \hat{b}_{k\mu}$,
and the system-reservoir interaction in the rotating frame is given by
$\hat{V}_\mu(t)=\sum_k\left[
g_{k\mu}e^{-i\omega_dt}\hat{b}_{k\mu}^\dagger \hat{A}_\mu
+g_{k\mu}^*e^{i\omega_dt}\hat{b}_{k\mu}\hat{A}_\mu^\dagger
\right]$.
The system-only rotating transformation leaves $H_b^\mu$ unchanged.
Therefore, the current is expressed as
\begin{align}
J_\mu(t){\equiv}\frac{d}{dt}\langle \hat{H}_b^\mu\rangle
=i\left\langle[\hat{V}_\mu(t),\hat{H}_b^\mu]\right\rangle.\label{eq:definition}
\end{align}
Using
$[\hat{H}_b^\mu,\hat{b}_{k\mu}^\dagger]=\omega_{k\mu}\hat{b}_{k\mu}^\dagger$
and
$[\hat{H}_b^\mu,\hat{b}_{k\mu}]=-\omega_{k\mu}\hat{b}_{k\mu}$,
we obtain the microscopic current operator
\begin{equation}
\hat{\mathcal{I}}_\mu(t)
=i\sum_k\omega_{k\mu}\left[
g_{k\mu}^*e^{i\omega_dt}\hat{b}_{k\mu}\hat{A}_\mu^\dagger
-g_{k\mu}e^{-i\omega_dt}\hat{b}_{k\mu}^\dagger \hat{A}_\mu
\right].\label{eq:microscopic-current}
\end{equation}
By introducing bosonic operators in the interaction picture
$\hat{B}_\mu(t)=\sum_k g_{k\mu}^*\hat{b}_{k\mu}e^{-i\omega_{k\mu}t}$,
$\hat{B}_{\mu,E}(t)=\sum_k\omega_{k\mu}g_{k\mu}^*
\hat{b}_{k\mu}e^{-i\omega_{k\mu}t}$,
and
$\hat{F}_\mu(t)=e^{-i\omega_dt}e^{i\hat{H}_{s,\rm{R}}t}\hat{A}_\mu e^{-i\hat{H}_{s,\rm{R}}t}$,
it is found that
$\hat{V}_{\mu,I}(t)=\hat{F}_\mu(t)\hat{B}_\mu^\dagger(t)
+\hat{F}_\mu^\dagger(t)\hat{B}_\mu(t)$
and
$\hat{\mathcal I}_{\mu,I}(t)=i\left[
\hat{F}_\mu^\dagger(t)\hat{B}_{\mu,E}(t)
-\hat{F}_\mu(t)\hat{B}_{\mu,E}^{\dagger}(t)
\right]$.

Based on the Born-Markov approximation, the first-order interaction-induced correction of the total density operator in the interaction picture is given by
\begin{equation}
\hat{\rho}_{\mathrm{tot},I}(t)
{=}\hat{\rho}_I(0)\otimes\hat{\rho}_b-i\int_0^\infty d\tau\,
\left[\hat{V}_I(t-\tau),\hat{\rho}_I(t)\otimes\hat{\rho}_b\right].\label{eq:correlation-correction}
\end{equation}
where the reduced quantum system density operator is described as
$\hat{\rho}_I(t)=e^{i\hat{H}_{s,\rm R}t}\hat{\rho}_{s,\rm R}(t)e^{-i\hat{H}_{s,\rm R}t}$
and the local equilibrium state of the thermal reservoirs is
$\hat{\rho}_b=\Pi_{\mu}(\exp(-\hat{H}^\mu_b/k_{\rm B}T_\mu)/Z_\mu)$
with $Z_\mu=\textrm{Tr}\{\exp(-\hat{H}^\mu_b/k_{\rm B}T_\mu)\}$.
Thus, the energy current into the $\mu$th reservoir can be expressed as
\begin{align}
J_\mu(t)={}&-i\int_0^\infty d\tau\,\operatorname{Tr}_{s,b}\Bigl\{
[\hat{\mathcal I}_{\mu,I}(t),\hat{V}_{\mu,I}(t-\tau)]\nonumber\\
&\qquad\times[\hat{\rho}_I(t)\otimes\hat{\rho}_b]\Bigr\}.
\label{eq:born-markov-current}
\end{align}
where cross terms involving different baths vanish for independent thermal
reservoirs.

Then, we introduce two relevant {energy-weighted bath correlations} 
\begin{subequations}
\begin{align}
C_{\mu,-}^{E}(\tau)
&=\left\langle \hat{B}_{\mu,E}(t)\hat{B}_\mu^\dagger(t-\tau)\right\rangle_b\nonumber\\
&=\sum_k\omega_{k\mu}|g_{k\mu}|^2
[n_\mu(\omega_{k\mu})+1]e^{-i\omega_{k\mu}\tau},\\
C_{\mu,+}^{E}(\tau)&=\left\langle \hat{B}_\mu^\dagger(t-\tau)\hat{B}_{\mu,E}(t)\right\rangle_b\nonumber\\
&=\sum_k\omega_{k\mu}|g_{k\mu}|^2
n_\mu(\omega_{k\mu})e^{-i\omega_{k\mu}\tau}.  
\end{align}
\end{subequations}
Both correlations have the phase $e^{-i\omega\tau}$,
and the factors $n_\mu(\omega_{k\mu})+1$ and $n_\mu(\omega_{k\mu})$ describe emission into and absorption from the bath, respectively.
Then, we define
\begin{align}
    {\overline {A}}_\mu(\tau)
=e^{i\omega_d\tau}e^{-i\hat{H}_{s,\rm R}\tau}\hat{A}_\mu e^{i\hat{H}_{s,\rm R}\tau}.\label{eq:evolved-operator}
\end{align}
Expanding the commutator in Eq.~\eqref{eq:born-markov-current}, performing
the bath trace, and returning to the rotating Schr\"odinger picture give
\begin{align}
J_\mu(t)={}&\int_0^\infty d\tau\,\operatorname{Tr}_s\Bigl\{\Bigl[
 C_{\mu,-}^{E}(\tau)\hat{A}_\mu^\dagger\overline A_\mu(\tau)\nonumber\\
&+C_{\mu,-}^{E*}(\tau)\overline A_\mu^\dagger(\tau)\hat{A}_\mu\nonumber\\
&-C_{\mu,+}^{E}(\tau)\overline A_\mu(\tau)\hat{A}_\mu^\dagger\nonumber\\
&-C_{\mu,+}^{E*}(\tau)\hat{A}_\mu\overline A_\mu^\dagger(\tau)
\Bigr]\hat{\rho}_{s,\rm R}(t)\Bigr\}.
\label{eq:four-terms}
\end{align}
The dressed operators that collect these integrals are
\begin{equation}
\hat{\mathcal {K}}_{\mu,\pm}
=\int_0^\infty d\tau\,
C_{\mu,\pm}^{E}(\tau)\overline A_\mu(\tau),\label{eq:kernel-operators}
\end{equation}
which simplify the expression for the energy current to
\begin{align}
J_\mu(t)=\operatorname{Tr}\Bigl[\Bigl(&
\hat{A}_\mu^\dagger\hat{\mathcal K}_{\mu,-}
+\hat{\mathcal K}_{\mu,-}^\dagger\hat{A}_\mu\nonumber\\
&-\hat{\mathcal K}_{\mu,+}\hat{A}_\mu^\dagger
-\hat{A}_\mu\hat{\mathcal K}_{\mu,+}^\dagger
\Bigr)\hat{\rho}_{s,\rm R}(t)\Bigr].
\label{eq:operator-current}
\end{align}

Furthermore, if we reexpress
$\hat{A}_\mu=\sum_{m,n}A^\mu_{mn}|\phi_m\rangle\langle\phi_n|$
with $H|\phi_m\rangle=E_m|\phi_m\rangle$,
equation~\eqref{eq:evolved-operator} gives
$\overline A_\mu(\tau)
=\sum_{m,n}A^\mu_{mn}
e^{iq_{mn}\tau}|\phi_m\rangle\langle\phi_n|$,
with $q_{mn}=\omega_d+E_n-E_m$.
Then, the dressed operator is
\begin{equation}
\hat{\mathcal K}_{\mu,\pm}
=\frac12\sum_{m,n}
q_{mn}\gamma_\mu(q_{mn})n_{\mu,\pm}(q_{mn})
A^\mu_{mn}|\phi_m\rangle\langle\phi_n|.
\label{eq:K}
\end{equation}
Substituting these operators into Eq.~\eqref{eq:operator-current} gives the concise expression of energy current
\begin{equation}
J_\mu(t)=2\mathrm{Re}\left\{\textrm{Tr}\left[
\left(\hat{A}_\mu^\dagger \hat{\mathcal K}_{\mu,-}
-\hat{A}_\mu \hat{\mathcal K}_{\mu,+}^{\dagger}\right)\hat{\rho}_{s,\rm R}(t)
\right]\right\}.\label{eq:general-current}
\end{equation}

\section{Steady-state energy flow in the nonequilibrium spin-boson model}\label{Append:C}

We start from the driven qubit Hamiltonian
\begin{align}
    \hat{H}_{\rm dQ1}=\Delta\hat{\sigma}_+\hat{\sigma}_--D(\hat{\sigma}_++\hat{\sigma}_-),
\end{align}
where the detuning is $\Delta=\varepsilon-\omega_d$.
Thus, the eigenvalues and eigenstates are straightforwardly obtained as
$E_{\pm}=(\Delta\pm\sqrt{\Delta^2+4D^2})/2$
and
$|\phi_+{\rangle}=\cos\frac{\theta}{2}|\uparrow{\rangle}-\sin\frac{\theta}{2}|\downarrow{\rangle}$
and
$|\phi_-{\rangle}=\sin\frac{\theta}{2}|\uparrow{\rangle}+\cos\frac{\theta}{2}|\downarrow{\rangle}$,
with $\tan\theta=2D/\Delta$.
After long-time evolution, it is found that for the driven spin-boson model,
the coherences of the reduced density operator in the eigenbasis of $\hat{H}_{\rm S,\rm R}$ are almost decoupled from the populations for weak system-bath interactions. Considering the microscopic energy exchange processes exhibited at Eqs.~(\ref{Dmu+}) and (\ref{Dmu-}), the incoherent transitions between two eigenstates are assisted by the driving frequency $\omega_d$. 
Consequently, the transition rates between $|\phi_{+}{\rangle}$ and $|\phi_{-}{\rangle}$ are given by
\begin{subequations}
    \begin{align}
    \Gamma^\mu_{\pm}(\omega_d+\Lambda)&=\pm\gamma_\mu(\omega_d+\Lambda)n_\mu(\pm(\omega_d+\Lambda))\cos^4\frac{\theta}{2},\\
      \Gamma^\mu_{\pm}(\omega_d-\Lambda)&=\pm\gamma_\mu(\omega_d-\Lambda)n_\mu(\pm(\omega_d-\Lambda))\sin^4\frac{\theta}{2},     
    \end{align}
\end{subequations}
with $\Lambda=\sqrt{\Delta^2+4D^2}$.
The transition rates between $|\phi_{\pm}{\rangle}$ and $|\phi_{\pm}{\rangle}$ are given by
\begin{align}
\Gamma^\mu_\pm(\omega_d)&=\pm\gamma_\mu(\omega_d)n_\mu(\pm\omega_d)\sin^2\frac{\theta}{2}\cos^2\frac{\theta}{2}.
\end{align}

The steady-state populations are obtained as
$P_+={\Gamma_+}/({\Gamma_++\Gamma_-})$ and
$P_-={\Gamma_-}/({\Gamma_++\Gamma_-})$,    
with collective rates $\Gamma_+=\sum_{\mu=l,r}[\Gamma^\mu_{+}(\omega_d+\Lambda)+\Gamma^\mu_{-}(\omega_d-\Lambda)]$
and
$\Gamma_-=\sum_{\mu=l,r}[\Gamma^\mu_{-}(\omega_d+\Lambda)+\Gamma^\mu_{+}(\omega_d-\Lambda)]$.
The steady-state energy flux into the $\mu$-th reservoir can be effectively described as
$J_\mu=\sum_{m,m^\prime}(\omega_d+E_{m^\prime{m}})[\Gamma^\mu_{-}(\omega_d+E_{m^\prime{m}})P_{m^\prime}-\Gamma^\mu_{+}(\omega_d+E_{m^\prime{m}})P_{m}]$,
with $P_m$ the steady-state population of the eigenstate $|\phi_m{\rangle}$. 
Therefore,
the analytical expressions for energy currents into the two reservoirs are
\begin{subequations}
\begin{align}
    J_l&=(\omega_d+\Lambda)[\Gamma^l_-(\omega_d+\Lambda)P_+-\Gamma^l_+(\omega_d+\Lambda)P_-]
    \nonumber\\
    &+(\omega_d-\Lambda)[\Gamma^l_-(\omega_d-\Lambda)P_--\Gamma^l_+(\omega_d-\Lambda)P_+]
    \nonumber\\
    &+\omega_d[\Gamma^l_-(\omega_d)-\Gamma^l_+(\omega_d)],\\
    J_r&=(\omega_d+\Lambda)[\Gamma^r_-(\omega_d+\Lambda)P_+-\Gamma^r_+(\omega_d+\Lambda)P_-]
    \nonumber\\
    &+(\omega_d-\Lambda)[\Gamma^r_-(\omega_d-\Lambda)P_--\Gamma^r_+(\omega_d-\Lambda)P_+]
    \nonumber\\
    &+\omega_d[\Gamma^r_-(\omega_d)-\Gamma^r_+(\omega_d)],\\    
    J_p&=-\{(\omega_d+\Lambda)[\Gamma_-(\omega_d+\Lambda)P_+-\Gamma_+(\omega_d+\Lambda)P_-]
    \nonumber\\
    &+(\omega_d-\Lambda)[\Gamma_-(\omega_d-\Lambda)P_--\Gamma_+(\omega_d-\Lambda)P_+]\nonumber\\
    &+\omega_d[\Gamma_-(\omega_d)-\Gamma_+(\omega_d)]\},
\end{align}
\end{subequations}
with the rates $\Gamma_{\pm}(\omega)=\sum_{\mu=l,r}\Gamma^\mu_{\pm}(\omega)$.

\end{document}